\documentclass[11pt,a4paper]{article}
\pdfoutput=1
\usepackage[applemac]{inputenc}
\usepackage{amsmath, amsthm}
\usepackage{jheppub}
\usepackage{multirow}
\usepackage{booktabs}
\usepackage{amsmath}
\usepackage{amsfonts}
\usepackage{amssymb}
\usepackage{graphicx}
\usepackage{indentfirst}
\usepackage{slashed}
\usepackage{mathrsfs}
\usepackage{tikz}
\usepackage{bm}

\usepackage[toc,page]{appendix}

\usepackage{etoolbox}

\appto\appendix{\addtocontents{toc}{\protect\setcounter{tocdepth}{1}}}

\appto\listoffigures{\addtocontents{lof}{\protect\setcounter{tocdepth}{1}}}
\appto\listoftables{\addtocontents{lot}{\protect\setcounter{tocdepth}{1}}}
\newcommand{\refE}[1]{eq.~(\ref{#1})}
\newcommand{\refF}[1]{fig.~\ref{#1}}

\newcommand{\refS}[1]{Section~\ref{#1}}

\newcommand{\abs}[1]{\left\lvert#1\right\rvert}

\newcommand{\res}{\text{Res}}
\newcommand{\vol}{\text{vol~}}

\newcommand{\Gram}{\text{Gram}}
\newcommand{\plainJ}{J}
\newcommand{\tildeJ}{\widetilde J}

\newcommand{\hypgeo}[4]{\,_2F_1\left(#1,#2;#3;#4\right)}

\newcommand{\rr}{r}

\newcommand{\bea}{\begin{eqnarray}}
\newcommand{\eea}{\end{eqnarray}}
\newcommand{\bean}{\begin{eqnarray*}}
\newcommand{\eean}{\end{eqnarray*}}

\def\half{\frac{1}{2}}

\def\abs#1{\left| #1\right|}

\def\det{\mathop{\rm det}}

\def\eps{\epsilon}

\def\ord{{\cal O}}

\def\Label#1{\label{#1}
  \smash{\hbox to0pt{\raise1ex\hbox{\tiny[#1]}\hss}}}

\def\beq{\begin{equation}}
\def\eeq{\end{equation}}
\def\bsp#1\esp{\begin{split}#1\end{split}}

\newcommand{\cC}{\begin{cal}C\end{cal}}

\newcommand{\cE}{\begin{cal}E\end{cal}}

\newcommand{\cutRes}{\mathcal{C}}

\newcommand{\plainK}{K}
\newcommand{\tildeK}{\widetilde{K}}

\def \LS{\textrm{L\hskip -1.3pt S}}

\renewcommand{\ln}{\log}

\theoremstyle{definition}

\preprint{CERN-TH-2017-033
\rightline{CP3-17-05}
\rightline{Edinburgh 2017/05}
\rightline{FR-PHENO-2017-001}
}

\title{Cuts from residues: the one-loop case}

\author[a]{Samuel Abreu,}
\author[b,c,d]{Ruth Britto,}
\author[e,f]{Claude Duhr,}
\author[g]{and Einan Gardi}

\affiliation[a]{Physikalisches Institut, Albert-Ludwigs-Universit\"at Freiburg, D-79104 Freiburg, Germany}
\affiliation[b]{School of Mathematics, Trinity College, Dublin 2, Ireland}
\affiliation[c]{Hamilton Mathematics Institute, Trinity College, Dublin 2, Ireland}
\affiliation[d]{Institute de Physique Th{\'e}orique, Universit\'e Paris Saclay, CEA, CNRS, F-91191 Gif-sur-Yvette, France}
\affiliation[e]{Theoretical Physics Department, CERN, Geneva, Switzerland}
\affiliation[f]{Center for Cosmology, Particle Physics and Phenomenology (CP3), Universit\'e Catholique de Louvain, 1348 Louvain-La-Neuve, Belgium}
\affiliation[g]{Higgs Centre for Theoretical Physics, 
School of Physics and Astronomy, \\
The University of Edinburgh, Edinburgh EH9 3FD, Scotland, UK}

\emailAdd{abreu.samuel@physik.uni-freiburg.de}
\emailAdd{ruth.britto@tcd.ie}
\emailAdd{claude.duhr@cern.ch}
\emailAdd{einan.gardi@ed.ac.uk}

\abstract{
Using the multivariate residue calculus of Leray, we give a precise definition of the notion of a cut Feynman integral in dimensional regularization, as a residue evaluated on the variety where some of the propagators are put on shell. These are naturally associated to Landau singularities of the first type. Focusing on the one-loop case, we give an explicit parametrization to compute such cut integrals, with which we study some of their properties and list explicit results for maximal and next-to-maximal cuts.
By analyzing homology groups, we show that cut integrals associated to Landau singularities of the second type are specific combinations of the usual cut integrals, and we obtain linear relations among different cuts of the same integral. We also show that all one-loop Feynman integrals and their cuts belong to the same class of functions, which can be written as parametric integrals.
}


\begin{document}
\maketitle

%%%%%%%%%%%%%%%%%%%%%%%%%%%%%%%%%%%%%%
%%%%%%%%%%%%%%%%%%%%%%%%%%%%%%%%%%%%%%
%%%%%%%%%%%%%%%%%%%%%%%%%%%%%%%%%%%%%%

%						MAIN TEXT

%%%%%%%%%%%%%%%%%%%%%%%%%%%%%%%%%%%%%%
%%%%%%%%%%%%%%%%%%%%%%%%%%%%%%%%%%%%%%
%%%%%%%%%%%%%%%%%%%%%%%%%%%%%%%%%%%%%%

% !TEX root = main-cuts.tex

\section{Introduction}

Precise predictions in perturbative quantum field theories require the calculation of loop integrals. The difficulty in evaluating these integrals greatly increases with the number of loops and the number of scales on which the integral depends. A better understanding of the analytic structure of these integrals has been fundamental in finding more efficient methods for their computation. In this paper, we will be concerned with one-loop integrals depending on an arbitrary number of scales.

It was realized in the early days of perturbative quantum field theories that cuts are an important tool to probe the analytic structure of Feynman integrals \cite{Cutkosky:1960sp,SMatrix,tHooft:1973pz}. Unitarity implies that Feynman integrals are multi-valued functions, and the cuts of Feynman integrals are related to the discontinuities. Singularities and branch cuts of Feynman integrals are classified by  the solutions to the {Landau conditions} \cite{Landau:1959fi}, a set of necessary conditions on the external data of an integral for a pinch singularity to occur. Modern unitarity methods build on this observation and, in a nutshell, use cuts to construct projectors onto a basis of master integrals~\cite{Bern:1997sc,Britto:2004nc,Forde:2007mi,Kosower:2011ty,CaronHuot:2012ab,Johansson:2012zv,Johansson:2013sda}. More recently, there has been a renewal in the interest in cut integrals in the study of integration-by-parts identities \cite{Sogaard:2014jla,Larsen:2015ped,Ita:2015tya} or differential equations~\cite{Remiddi:2016gno,Primo:2016ebd,Frellesvig:2017aai,Zeng:2017ipr} satisfied by Feynman integrals, and in applications of the solutions to Landau conditions \cite{Dennen:2015bet,Dennen:2016mdk}.

Loosely speaking, cut integrals are computed by replacing a subset of the propagators that are called \emph{cut} by Dirac-$\delta$ functions, and the integral is then evaluated under these constraints. However, if one wishes to study the analytic structure of Feynman integrals, one must be more precise in the definition of cuts. Such precise definitions exist for certain types of cuts. For instance, one can consider so-called unitarity cuts \cite{tHooft:1973pz,Veltman:1994wz,Remiddi:1981hn}, which select a particular external channel, or iterated unitarity cuts~\cite{Mandelstam:1959bc,Ball:1991bs,Abreu:2014cla}  which select different channels. When propagators are massive, one can also consider single-propagator cuts~\cite{Abreu:2015zaa}. These precise definitions are tailored to compute discontinuities in the variables identified with external channels or internal masses. However, they do not exhaust the complete set of cuts one might wish to compute. 
For example, it is well known that the propagators of a massless four-point one-loop integral have no common zero on the real axis~\cite{Britto:2004nc,Kosower:2011ty}.  Indeed, if pole of any cut propagator does not lie inside the integration region, then the cut would be zero according to the definitions above. 
This has led to the idea that cuts should be computed via residues, i.e., by deforming the integration contour such that it encircles the poles of the cut propagators~\cite{Cachazo:2008vp,ArkaniHamed:2008gz,Kosower:2011ty}. While this procedure is very clear in the case of integrals where the number of dimensions matches the number of propagators, it is often not entirely clear what the correct integration contour is in cases where not all integration can be done using the residue theorem.

The aim of this paper is to study one-loop cut integrals and to give a precise definition of cut Feynman integrals as residues integrated over a well-defined contour in dimensional regularization. The motivation to study these objects is mostly driven by a desire to improve our understanding of the analytic structure of loop integrals, in particular in the light of novel mathematical developments and an improved understanding of the functions that appear in loop computations. For example, it was shown in concrete examples \cite{Abreu:2014cla,Abreu:2015zaa} that the coproduct of loop integrals can be cast in a form such that the rightmost entries are cut integrals. A complete understanding of this observation, however, requires a rigorous definition of the relevant integration contours and of how to evaluate the cut integrals, including for non-integer dimensions in order to work in dimensional regularization. To our knowledge, this information is not hitherto available in the literature. For example, while it is clear how to evaluate the quadruple cut of a box integral, it is less obvious how to precisely determine the correct contour and evaluate a triple cut, where one still needs to perform one integration. Moreover, while the single and double cuts of a box integral have a clear interpretation in terms of discontinuities in masses and external channels, it is less clear how to interpret the discontinuity of the box integral computed by the triple and quadruple cuts. Even less is known about how to answer these questions in the case of pinch singularities at infinite loop momentum, the so-called Landau singularities of the second type~\cite{SMatrix,SecondType1,SecondType2}.

In this paper we close this gap in the literature and perform the first rigorous study of cut integrals in dimensional regularization. We focus  on cut integrals at one loop, though we expect that many of the concepts we introduce in this paper are generic and will carry through to higher loops. In fact, many of these concepts have been introduced into the mathematical physics literature in the 60s~\cite{Cutkosky:1960sp,SMatrix,PhamCompact,Teplitz,PhamBook} (albeit without the machinery of dimensional regularization), but they have since slipped into oblivion. The cornerstone of our approach to cut integrals is the multivariate residue calculus of Leray~\cite{Leray}.
In this setup, the integrand is modified by evaluating its residues at the poles of the cut propagators, and this new integrand is integrated over the \emph{vanishing sphere}. Through a generalization of the residue theorem, the cut integral can also be written as an integral over the \emph{vanishing cycle}, in which case the integrand is the same as for the uncut integral. Moreover, cuts are intimately connected to discontinuities through the Picard-Lefschetz theorem, which relates the change of the integration contour under analytic continuation to integrals of residues over the vanishing spheres. The study of the vanishing cycles naturally leads to the study of the homology group associated to one-loop integrals, which is the right language to discuss the different inequivalent integration contours for one-loop integrals. 

By choosing concrete parametrizations of the loop momenta, we can use our definition of cut Feynman integrals to compute them explicitly. We find they agree with classic unitarity cuts whenever the latter are well defined.
The one-loop framework we provide allows us to carry out the integral to all orders in dimensional regularization for maximal and next-to-maximal cuts in full generality, and to understand which cuts vanish identically. Through compactification of one-loop integrals, we are able to combine cut and uncut integrals into the same class of parametric integrals. We find this framework suitable for studying connections to (iterated) discontinuities. Furthermore, our analysis of homology leads to classes of linear relations among different cuts of the same integral.

This paper is organized as follows: In Section~\ref{Sec:one-loop} we give a short review of one-loop integrals and we set up our notation and conventions.
In Section~\ref{sec:residues} we present our definition of cut integrals via Leray's multivariate residue calculus, and in Section~\ref{sec:explicitRes} we give concrete results for certain classes of cut integrals, including vanishing cuts and maximal and next-to-maximal cuts. In Section~\ref{sec:secondType} we discuss the homology groups associated to one-loop integrals and use them to define cut integrals associated to singularities of the second type. In Section~\ref{sec:loop-cut-duality}, we introduce a class of parametric integrals that allows us to compute both cut and uncut one-loop integrals. In Section~\ref{sec:cuts_&_discontinuities} we review the Picard-Lefschetz theorem and how it connects to the concepts  of discontinuities  and leading singularities in the physics literature. In Section~\ref{sec:ibps} we discuss linear relations among cut integrals, and in Section~\ref{sec:conclusions} we draw our conclusions. We include several appendices where we present technical details that are omitted throughout the main text.

% !TEX root = main-cuts.tex

\section{One-loop integrals}
\label{Sec:one-loop}

Consider a one-loop Feynman integral with $n$ propagators in $D=d-2\epsilon$ dimensions, where $d$ is an even integer. One-loop integrals with numerators and/or higher powers of the propagators can always be reduced to a linear combination of integrals where all propagators are raised to unit powers.\footnote{We are grateful to Roman Lee and Volodya Smirnov for correspondence on how to prove this statement rigorously to all orders in $\eps$ in dimensional regularization.}  We therefore only concentrate on integrals of the following type,
\begin{equation}\label{integral}
I_n^D\left(\left\{p_i\cdot p_k\right\};\left\{m_{j}^2\right\}\right)=\frac{e^{\gamma_{E}\epsilon}}{i\pi^{{D}/{2}}}\int d^Dk\prod^{n}_{j=1}\frac{1}{\left(k-q_j\right)^2-m_j^2+i0}\,,
\end{equation}
where $\gamma_E = -\Gamma'(1)$ is the Euler-Mascheroni constant. The external momenta $p_i$ satisfy momentum conservation, which we write in the form
\beq
\sum_{i=1}^{n}p_i=0\,.
\eeq
The $m_j$ are the internal masses associated respectively to the propagators carrying momentum $k-q_j$, and   the $q_j$ are combinations of the external momenta $p_i$,
\begin{equation}
q_j=\sum_{i=1}^{n} c_{ji}\, p_i\,,\qquad c_{ji}\in\{-1,0,1\}.
\end{equation}
We define the loop momentum $k$ as the momentum carried by the propagator labelled by $1$,
so that $q_{1}$ is the zero vector, $q_{1}=\mathbf{0}_{D}$.
In general, we will not explicitly write the variables on which $I_n^D$ depends and suppress the superscript $D$.

Since all one-loop integrals with numerators and/or higher powers of the propagators can be reduced to integrals of the type~\eqref{integral}, these integrals form a basis of all one-loop integrals.\footnote{Some two- and three-point integrals of the type~\eqref{integral} with specific kinematic configurations are reducible to one- and two-point functions, so the set of all integrals of the type~\eqref{integral} is strictly speaking over-complete.} Integrals in different space-time dimensions are related through dimensional-shift identities~\cite{Bern:1992em,Tarasov:1996br,Lee:2009dh}, and so it is sufficient to consider basis integrals in a fixed number of dimensions. 
It is, however, often convenient to choose integrals with different numbers of external legs to lie in different dimensions, and in this paper we consider the 
following set of integrals,
\beq\label{eq:tildeJ_n}
\tildeJ_n(\left\{q_i\cdot q_k\right\};\left\{m_{j}^2\right\};\eps) \equiv I_n^{D_n}(\left\{q_i\cdot q_k\right\};\left\{m_{j}^2\right\})\,,
\eeq
where
\beq\label{eq:defDn}
D_n=\left\{\begin{array}{ll}
n-2\eps\,, & \textrm{ if } n \textrm{ even}\,,\\
n+1-2\eps\,, & \textrm{ if } n \textrm{ odd}\,.
\end{array}\right.
\eeq
The functions $\tildeJ_n$ form a basis for the vector space spanned by all one-loop Feynman integrals in $D=d-2\eps$ dimensions, $d$ an even integer. The advantage of this set over a basis where all integrals lie in the same dimension is that, conjecturally, all the elements of this basis are, order by order in $\eps$, polylogarithmic functions of uniform transcendental weight $\lceil n/2\rceil$ (we consider $\eps$ to have weight $-1$), where $\lceil .\rceil$ is the ceiling function which gives the smallest integer greater than (or equal to) its argument.
Although this statement has only been proved for all dual-conformally-invariant integrals $\tildeJ_n$ with $n$ even \cite{Spradlin:2011wp}, there is strong indication that it holds in general.

It is clear that one-loop Feynman integrals are invariant under dihedral transformations generated by rotations and reflections,
\beq
(q_i,m_i^2) \to (q_{i+1},m_{i+1}^2) {\rm~~and~~} (q_i,m_i^2) \to (q_{n-i+1},m^2_{n-i+1})\,,
\eeq
and all indices are understood modulo $n$.
Since every one-loop graph is planar, we can view the variables $k, q_1,\ldots,q_{n}$ as the dual momentum coordinates of the one-loop graph defining the integral $I_n$: each of these variables can be associated to a face of the original graph, or equivalently to a vertex of the dual graph. The dual graph makes apparent an enhanced symmetry of our basis of one-loop Feynman integrals:  the dual representation is manifestly symmetric under any permutation of the propagators.

In the remainder of this paper we define a cut integral as a variant of a one-loop Feynman integral, where some of the propagators are put on their mass shells. More precisely, a cut integral corresponds to the original Feynman integral evaluated on a contour that encircles some of the poles of the propagators. This contour integral can be evaluated in terms of residues.

% !TEX root = main-cuts.tex
\section{Cuts and residues}
\label{sec:residues}

Discontinuities are closely related to residues. In order to define cut integrals and their relation to discontinuities of Feynman integrals, we need a generalization of the usual residue calculus to the multivariable case. We start by reviewing the multivariate residue calculus of Leray~\cite{Leray}.  We then define cut integrals in terms of multivariate residues and discuss the geometric interpretation of the contours of integration.

\subsection{Multivariate residues}\label{sec:mvresidues}

Leray's multivariate residues are most conveniently defined in the language of differential forms. Consider a space $X$ and an irreducible subvariety $S$ of $X$ defined by the equation $s(z)=0$, where $z$ denotes a set of coordinates on $X$. If $\omega$ is a differential $k$-form defined on the complement $X- S$ of $S$, then we say that $\omega$ has a \emph{pole of order $n$ on $S$} if $s^n\omega$ can be extended to a regular form on all of $X$ that is nonvanishing on $S$. One can show that if $\omega$ has a pole of order $n$ on $S$, then there are differential forms $\psi$ and $\theta$ such that
\beq\label{eq:to_residue}
\omega = \frac{ds}{s^n}\wedge \psi + \theta\,,
\eeq
where $\psi$ is regular and nonvanishing on $S$, and $\theta$ has a pole of order at most $n-1$ on $S$. In the special case of a simple pole, $n=1$, the residue of $\omega$ on $S$ is defined as the restriction of $\psi$ to the subvariety $S$,
\beq
\textrm{Res}_S[\omega] = \psi_{|S}\,.
\eeq
The definition of the residue can be extended to poles of higher order using the Leibniz rule. Indeed, if $\omega_n$ has a pole of order $n$, we have
\beq
\omega_n = \frac{ds}{s^n}\wedge \psi + \theta = d\left(-\frac{\psi}{(n -1)\,s^{n-1}}\right) + \frac{d\psi}{(n -1)\,s^{n-1}} + \theta\,.
\eeq
We see that, up to an exact form (i.e., up to a total derivative), $\omega_n$ is equivalent to the form $\omega_{n-1} \equiv \frac{d\psi}{(n -1)\,s^{n-1}} + \theta$, which has a pole of order at most $n-1$. Iterating this procedure, we see that every form is equivalent (up to an exact form) to a form $\omega_{1}$ with at most a simple pole. The residue of $\omega_n$ is then defined to be equal to the residue of $\omega_1$,
\beq
\textrm{Res}_S[\omega_n] \equiv \textrm{Res}_S[\omega_1]\,.
\eeq
Technically speaking, the previous argument shows that the cohomology class of every form contains a form with at most a simple pole on $S$, and the residue map is well defined on cohomology classes. In other words, if $H^k_{dR}$ denotes the $k$-th de Rham cohomology group, then we may interpret the residue as a map $\textrm{Res}_S : H^k_{dR}(X- S)\to H^{k-1}_{dR}(S)$.

The previous definition generalizes the notion of residue from complex analysis. Indeed, if $X=\mathbb{C}$, then an irreducible subvariety $S$ necessarily has the form $s(z) = z-a=0$, i.e., it is an isolated point in the complex plane. Consider the one-form 
\beq\label{eq:one-form}
\omega_n = \frac{g(z)\,dz}{(z-a)^n}\,,
\eeq
where $g$ is holomorphic and nonvanishing at $z=a$. Using the Leibniz rule, it is easy to check that 
\beq
\omega_{n-l} = \frac{g^{(l)}(z)\,dz}{(n-1)\ldots (n-l)\,(z-a)^{n-l}}\,,
\eeq
and so the residue is the zero-form
\beq
\textrm{Res}_S[\omega_n] = \textrm{Res}_S[\omega_1] = \frac{g^{(n-1)}(a)}{(n-1)!}\,,
\eeq
in agreement with the usual residue calculus. 

At this point, this definition of multivariate residues is a property only of differential forms, and it does not make reference to any contour integration. We now discuss the interplay of multivariate residues and contour integrals, in particular the generalization of the residue theorem to the multivariate case. We first need to define the equivalent of an integration contour that encircles the singular surface $S$ in the case where $S$ is not just a single point, but a variety of codimension 1. Consider a $k$-cycle $\sigma\subset S$. Since $S$ has complex codimension 1 (i.e., real codimension 2), to each point $P\in S$ we can associate a small circle in the complex plane `transverse' to $S$ and centered on $P$. If we carry out this construction for every point of the $k$-cycle $\sigma$, we obtain a $(k+1)$-cycle $\delta\sigma$, called the \emph{tubular neighborhood}, which `wraps around' the $k$-cycle $\sigma$. By construction, $\delta\sigma$ does not intersect $\sigma$. The linear operator $\delta$ which assigns to a $k$-cycle its tubular neighborhood is called the \emph{Leray coboundary}. 

The tubular neighborhood and the Leray coboundary provide a generalization of the residue theorem to the multivariate case. More precisely, if $\omega$ is a $(k+1)$ form on $X- S$ and $\sigma$ is a $k$-cycle in $S$, then we have~\cite{Leray}
\beq\label{eq:residue_thm}
\int_{\delta\sigma}\omega = 2\pi i\,\int_{\sigma}\textrm{Res}_S[\omega]\,.
\eeq
The right-hand side is well defined because the residue is regular on the singular surface $S$, and the left-hand side because $\delta\sigma$ is a $(k+1)$-cycle on $X- S$.

Let us illustrate that eq.~\eqref{eq:residue_thm} reduces to the usual residue theorem in the case where $X=\mathbb{C}$, the singular surface $S$ is the isolated point defined by $s(z)=z-a=0$, and we consider the one-form $\omega_n$ defined in eq.~\eqref{eq:one-form}. Since $S$ is an isolated point, it contains a single 0-cycle $\sigma$, which is the point $a$ itself. The tubular neighborhood of a point is a small circle around this point. Hence, we obtain
\beq
\int_{\delta\sigma}\omega_n = \oint  \frac{g(z)\,dz}{(z-a)^n} = 2\pi i\,\frac{g^{(n-1)}(a)}{(n-1)!} = 2\pi i\,\int_{z=a}\textrm{Res}_S[\omega_n]\,.
\eeq

We conclude our discussion of the generalized residue theorem with a comment on the interpretation of the residue map and the Leray coboundary. One can show that the Leray coboundary of a cycle is a cycle and that of a boundary a boundary. Therefore $\delta$ is well-defined on (singular) homology classes, i.e., it defines a map $\delta: H_k(S) \to H_{k+1}(X- S)$. It is known from de Rham's theorem that the (complexified) de Rham cohomology and singular homology groups are dual to each other, where the duality is expressed by the bilinear form defined by the integration of differential forms over cycles,
\beq
\langle .|.\rangle : H_k \times H^k_{dR} \to \mathbb{C}\,; \quad\,(\sigma, \omega) \mapsto \langle \sigma|\omega\rangle \equiv \int_{\sigma}\omega\,.
\eeq
In this context the Leray coboundary and residue maps can be understood as dual to each other, because we can use this bilinear form to write the residue theorem as
\beq
\langle\delta\sigma | \omega\rangle = 2\pi i\,\langle \sigma|\textrm{Res}_S[\omega]\rangle\,.
\eeq

So far, we have only considered the situation where $\omega$ has a pole on a single subvariety $S$. In the following we also need to consider the case where $\omega$ has poles on a family of subvarieties $S_i$, $1\le i\le m$, defined by the equations $s_i(z)=0$. We only discuss the case $m=2$, as the generalization to  general $m$ is straightforward. Moreover, it is sufficient to assume that $\omega$ has simple poles along each singular surface $S_i$ (because otherwise we can replace it up to an exact form by a form that only has simple poles). Iterating the previous discussion, we see that we can write $\omega$ in the form
\beq
\omega = \frac{ds_1}{s_1}\wedge \frac{ds_2}{s_2}\wedge \psi_{12} + \frac{ds_1}{s_1}\wedge \psi_{1} + \frac{ds_2}{s_2}\wedge \psi_{2} + \theta\,,
\eeq
where $\psi_{1}$, $\psi_{2}$ are regular on each $S_i$, $\psi_{12}$ on $S_1\cap S_2$ and $\theta$ everywhere. The \emph{composed residue} of $\omega$ on $S_1$ and $S_2$ is defined as the restriction of $\psi_{12}$ to $S_1\cap S_2$,
\beq
\textrm{Res}_{S_1S_2}[\omega] = \psi_{12|S_1\cap S_2}\,.
\eeq
The definition makes it clear that the composed residue is antisymmetric in the order of the singular surfaces: $\textrm{Res}_{S_1S_2} = -\textrm{Res}_{S_2S_1}$. Composed residues at poles of higher order are defined in the obvious way. In particular, the composed residue map is well defined on cohomology classes, and we obtain a map $\textrm{Res}_{S_1S_2} : H^k_{dR}(X- (S_1\cup S_2))\to H^{k-2}_{dR}(S_1\cap S_2)$.

A special case of the previous definition is the residue at a global pole. For $\textrm{dim}\, X = n$, consider the differential $n$-form
\beq
\omega = \frac{h(z)\,d^nz}{s_1(z)\dots s_n(z)}\,.
\eeq
For simplicity, we only discuss the case of simple poles. If we change variables to $y_i = s_i(z)$, we find
\beq
\omega = \frac{h(y)\,d^ny}{J(y)\,y_1\ldots y_n}\,,
\eeq
where $J(y)$ denotes the jacobian. It is now easy to see that the composed residue agrees with the value of the residue at the global pole (up to the sign coming from the ordering of the singular surfaces),
\beq
\textrm{Res}_{S_1\ldots S_n}[\omega] = \pm\frac{h(0)}{J(0)}\,.
\eeq

We can generalize the residue theorem~\eqref{eq:residue_thm} to the situation where we have multiple singular surfaces. In the special case of two singular surfaces, it reads
\beq
\int_{\delta_{S_1S_2}\sigma}\omega = (2\pi i)^2\,\int_{\sigma}\textrm{Res}_{S_2S_1}[\omega]\,,
\eeq
where $\omega$ is a $(k+2)$-form on $X- (S_1\cup S_2)$ and $\sigma$ is a $k$-cycle in $S_1\cap S_2$. The iterated Leray coboundary is defined in the obvious way,
\beq
\delta_{S_1S_2} \equiv \delta_{S_1} \delta_{S_2}\,,
\eeq
where $\delta_{S_i}$ denotes the Leray coboundary associated to the singular surface $S_i$. The composition of Leray coboundaries is antisymmetric in order to compensate for the antisymmetry of the composed residue map:
\beq\label{eq:Leray_antisymmetry}
\delta_{S_1S_2} = -\delta_{S_2S_1}\,.
\eeq

% !TEX root = main-cuts.tex

\subsection{Definition of cut integrals}
\label{sec:def_cuts}

Let us now turn to our definition of one-loop cut integrals. 
Let $C$ denote a subset of propagators that are called \emph{cut}, while the remaining propagators are called \emph{uncut}.  
Following the usual approach in the physics literature, we want to define a cut integral as the original loop integral where the contour has been replaced by a contour $\Gamma_C$ which encircles the poles of the cut propagators (and no other poles). As a consequence, we can take residues at the locations of the poles of the cut propagators. In this section we give a rigorous definition of this procedure using the concepts from multivariate residue calculus reviewed in the previous section. In a nutshell, to every integrand (i.e. differential form) of a Feynman integral, we can associate a new integrand by acting with the composed residue map corresponding to the singular surfaces where the propagators in $C$ are on shell. The resulting integrand can be naturally integrated over a cycle which corresponds to the intersection of the singular surfaces. Using the generalized residue theorem~\eqref{eq:residue_thm}, we can relate the integral over the locus where the propagators are on shell to the original loop integral over a deformed contour $\Gamma_C$, whose homology class is defined unambiguously using the iterated Leray coboundary. In the remainder of this section we discuss all these steps in detail. We focus on one-loop integrals, although many of the concepts easily generalize beyond one loop.

Let us start by defining the residues of a one-loop integral. We know that the residue map acts on differential forms, and it is therefore convenient to cast eq.~\eqref{integral} in the form
\beq
I_n^D = \int\omega_n^D\,,
\eeq
where we define the differential form
\beq
\omega_n^D = \frac{e^{\gamma_{E}\epsilon}}{i\pi^{{D}/{2}}} \frac{d^Dk}{D_1\ldots D_{n}}\,,
\end{equation}
with $D_j = \left(k-q_{j}\right)^2-m_{j}^2+i0$. The total symmetry of one-loop integrals implies that we can assume, without loss of generality, that the set $C$ of cut propagators is $C=[c]$ with
\begin{equation}
	[c]\equiv\{1,\ldots,{c}\}\,.
\end{equation}
We then define $\cE_C$ to be the linear subspace spanned by the vectors $q_i$, $i\in C$.

Our first goal is to compute the composed residue $\textrm{Res}_C[\omega_n^D] \equiv \textrm{Res}_{S_1\ldots S_{c}}[\omega_n^D]$, where $S_i$ denotes the hypersurface where the $i$-th propagator is on shell, $D_i=0$. Note that $\textrm{Res}_C[\omega_n^D]$ is only defined up to a sign, corresponding to the ordering of the singular surfaces. We only discuss the case of propagators with unit powers, but the generalization to arbitrary integer powers is straightforward (one simply applies the result for poles of higher order quoted in the previous section). 

In order to evaluate the residues, we need to write $\omega_n^D$ in a form which mimics eq.~\eqref{eq:to_residue}. This can be achieved by changing a subset of integration variables to be the cut propagators~\cite{Cutkosky:1960sp,Baikov:1996iu,Lee:2010wea,Grozin:2011mt}. We start by decomposing the loop momentum as $k=k_{\parallel} + k_{\bot}$, where $k_{\parallel}$ denotes the projection of $k$ onto the subspace $\cE_C$. In this subspace we change integration variables to the scalar products $k\cdot (q_i-q_{\ast})$, $i\in C\setminus\{\ast\}$, where $*$ denotes any particular element of $C$ (for example the element with the lowest index). We introduce polar coordinates in the transverse space, and we change variables from $k_{\bot}^2$ to $k^2$. 
This new parametrization has the advantage that all the cut propagators are linear in the new variables $k^2$ and $k\cdot q_i$, $i\in C\setminus\{\ast\}$, and so we can easily change variables to the cut propagators $D_i$, $i\in C$. At the end of this procedure, the differential form $\omega_n^D$ can be written as
\beq\label{eq:Baikov}
\omega_n^D= \frac{2^{-c}\,e^{\gamma_E\eps}}{\sqrt{\mu^c\,H_C}}\,\left(\mu\,\frac{H_C}{\Gram_C}\right)^{(D-c)/2}\,\frac{d\Omega_{D-c}}{i\pi^{D/2}}\,
\left(\prod_{j \notin C}\frac{1}{D_j}\right)\,\left(\prod_{j\in C}\frac{dD_j}{D_j}\right)\,,
\eeq
where $\mu=+1$ $(-1)$ in Euclidean (Minkowski) signature,  $d\Omega_{D-c}$ denotes the integration measure on the $(D-c)$-sphere $S_\bot\simeq S^{D-c}$ in the $(D-c+1)$-dimensional transverse space, and we have introduced the Gram determinants\footnote{The signs of the determinants depend on the order of the elements of $C$. From here on, we assume that a definite order has been chosen. }
\begin{align}
\label{eq:gramHsubsetC}
H_C &= \det\left((k-q_i) \cdot(k-q_j) \right)_{i,j \in C}\, ,\\
\label{eq:gramGsubsetC}
\Gram_C &= \det\left((q_i-q_*) \cdot(q_j-q_*) \right)_{i,j \in C\setminus\{\ast\}}\,.
\end{align}
A detailed proof of eq.~\eqref{eq:Baikov} is presented in Appendix~\ref{app:Baikov}.
 
Equation~\eqref{eq:Baikov} is precisely the form that we need to compute the composed residue of $\omega_n^D$, and we immediately find
\beq
\textrm{Res}_C[\omega_n^D] = 2^{-c}\,e^{\gamma_E\eps}\,\frac{d\Omega_{D-c}}{i\pi^{D/2}}\,\left[\frac{1}{\sqrt{\mu^c\,H_C}}\,\left(\mu\,\frac{H_C}{\Gram_C}\right)^{(D-c)/2}\,\left(\prod_{j\notin C}\frac{1}{D_j}\right)\right]_C\,,
\eeq
where the notation $[.]_C$ indicates that the expression inside square brackets should be evaluated on the locus where the cut propagators vanish. We can further simplify this expression by noting that the Gram determinant $[H_C]_C$ can be written in a more familiar form, manifestly independent of $k$. Indeed, if $(k-q_i)^2=m_i^2$ and $(k-q_j)^2=m_j^2$, then
\beq\bsp
2(k-q_i) \cdot(k-q_j)
% &= 2 (k^E)^2 - 2 k^E \cdot q^E_i- 2 k^E \cdot q^E_j + 2 q^E_i \cdot q^E_j \\
&= (k-q_i)^2 + (k-q_j)^2 -  (q_i-q_j)^2 \\
&= m_i^2 + m_j^2 -  (q_i-q_j)^2\,.
\esp\eeq
and therefore
\beq\bsp
[H_C]_C = Y_C\,,
\esp\eeq
where $Y_C$ is the \emph{modified Cayley determinant}, defined by
\beq\bsp
\label{eq:mcayleyYsubsetC}
Y_C = \det \left(\frac{1}{2}(-(q_i-q_j)^2+m_i^2+m_j^2)\right)_{ i,j \in C}\,.
\esp\eeq
Hence, we can write the residue in the form
\beq\bsp
\label{eq:residue-def}
\textrm{Res}_C[\omega_n^D] = \frac{2^{-c}\,e^{\gamma_E\eps}}{\sqrt{\mu^c\,Y_C}}\,\left(\mu\,\frac{Y_C}{\Gram_C}\right)^{(D-c)/2}
\frac{d\Omega_{D-c+1}}{i\pi^{D/2}}\,\left[\prod_{j \notin C}\frac{1}{(k-q_j)^2-m_j^2}\right]_{C}\,.
\esp\eeq
We recall that the residue is only defined up to a sign that varies according to the ordering of the singular surfaces.

Let us now turn to our definition of cut {integrals}.
The residue in eq.~\eqref{eq:residue-def} is a differential form on the $(D-c)$-sphere $S_{\bot}$ in the transverse space. It is therefore natural to define a one-loop cut integral as 
\begin{align}\label{eq:cutdef-det}
\cC_C&I_n  \equiv  \,(2\pi i)^{\lfloor c/2\rfloor}\,\int_{S_{\bot}}\textrm{Res}_C[\omega_n^D]\mod i\pi\\
\nonumber
 &\,= 2^{-c}\,\frac{(2\pi i)^{\lfloor {c}/{2}\rfloor}\,e^{\gamma_E\eps}}{\sqrt{\mu^c\,Y_C}}\,\left(\mu\,\frac{Y_C}{\Gram_C}\right)^{(D-c)/2}\!\!\!
%\\ &\,\times
\int_{S_\bot}\frac{d\Omega_{D-c}}{i\pi^{D/2}}\left[\prod_{j \notin C}\frac{1}{(k-q_j)^2-m_j^2}\right]_{C}\!\!\!\!\mod i\pi\,,
\end{align}
where the normalization factor $(2\pi i)^{\lfloor c/2\rfloor}$ has been introduced to allow us to remove the leading power of $i\pi$. Just like the residue form in eq.~\eqref{eq:residue-def}, our cut integrals are only defined up to a sign. In addition, depending on the kinematic point, the ratio of determinants raised to non-integer powers, as well as the integral over the sphere, may develop imaginary parts for which a prescription needs to be defined. For the purpose of this paper, we are not concerned with the value of this region-dependent imaginary part, which means that the value of $\cC_CI_n$ is defined only up to branch cuts. This is why we assume from now on that our definition~\eqref{eq:cutdef-det} is only valid modulo $i\pi$.

\subsection{The integration contour}
Let us discuss in more detail the choice of the integration contour $S_{\bot}$ in eq.~\eqref{eq:cutdef-det}. We will show that the sphere $S_{\bot}$ can be identified with the intersection of the singular surfaces, $S_C \equiv \bigcap_{j\in C}S_j$.  Throughout this section we work in real Euclidean kinematics, because it makes the geometric intuition more transparent. In other words, we perform an analytic continuation to Euclidean momenta
\beq\bsp
k^E=\left(-ik_0,k_1,\ldots,k_{D-1}\right) {\rm~~and~~} q_j^E&=\left(-iq_{j0},q_{j1},\ldots,q_{j(D-1)}\right)\,.
\label{eq:pre-euc-param}
\esp\eeq
The Euclidean vectors satisfy
\begin{equation}
\left(k^E\right)^2=-k^2\,,\qquad \left(q_j^E\right)^2=-q_j^2\,,\qquad k^E\cdot q_j^E=-k\cdot q_j\,.
\end{equation}
The transition to Euclidean kinematics induces a sign change in both the Gram and modified Cayley determinants ($(-1)^c$ in (\ref{eq:gramHsubsetC}) and $(-1)^{c-1}$ in (\ref{eq:gramGsubsetC})), which amounts to changing the sign of $\mu$ in eq.~\eqref{eq:Baikov}.

In order to define the residue, we have decomposed the space $\mathbb{R}^D$ in which the loop momentum $k^E$ lives as
\beq
\mathbb{R}^D = \cE_C \times \mathbb{R}_+ \times S_{\bot}\,,
\eeq
where the subspace $\cE_C \times \mathbb{R}_+$ is parametrized by $(k^E)^2$ and $k^E\cdot q_j^E$, $j\in \{2,\ldots,c\}$ (recall that in our parametrization we take $q_1=q^E_1=\mathbf{0}_D$). 
The on-shell constraints,
\bea\label{eq:sing_surf}
(k^E-q^E_j)^2+m_j^2 = 0, \qquad  \forall  j \in C\,,
\eea
fix all the components of $k^E$ in the subspace $\cE_C \times \mathbb{R}_+$, but the degrees of freedom of $k^E$ parametrized by the sphere $S_{\bot}$ are unconstrained. Moreover, the on-shell constraints~\eqref{eq:sing_surf} allow us to identify the singular surfaces $S_j$ as $(D-1)$-spheres.
In other words, the remaining components of $k^E$ are constrained to lie on the intersection of the singular surfaces $S_j$.  Since $S_C$ and $S_\bot$ have the same dimension, the intersection $S_C$ must coincide with the sphere $S_{\bot}$. 

Let us explain more explicitly why the intersection $S_C$ is a sphere.
We can subtract one of the equations in eq.~\eqref{eq:sing_surf} from the $(c-1)$ remaining equations, and we obtain
\beq
\left\{\begin{array}{rll}
(k^E-q^E_{i})^2 &= -m_{i}^2\,, &\qquad i\in C\\
2 k^E \cdot (q^E_{i} - q^E_{j}) &= (q^E_{i})^2 + m_{i}^2-(q^E_{j})^2 - m_{j}^2   \,, &\qquad j\in C\,, \,j\neq i\,.
\end{array}\right.
\eeq
We see that the intersection of $c$ spheres is equivalent to the intersection of a single sphere with $(c-1)$ hyperplanes. In order to conclude that this intersection is a sphere, it is sufficient to note that the intersection of a sphere with \emph{one} hyperplane is again a sphere of one dimension less. It follows inductively that the total intersection $S_C$ is a $(D-c)$-sphere.  

A recurrent theme in the physics literature states that  cuts correspond to integrals of the original integrand $\omega_n^D$ over a deformed contour $\Gamma_C$ that encircles the poles of the cut propagators. We can now make this statement concrete, and we explicitly construct this integration contour. 
Indeed, from $S_{\bot}=S_C$ it follows that $S_{\bot}$ is a cycle in $S_C$, and so the generalized residue theorem~\eqref{eq:residue_thm} applied to the cut integral in eq.~\eqref{eq:cutdef-det} gives
\beq\label{eq:cut_sphere}
\cC_CI_n = (2\pi i)^{\lfloor c/2\rfloor}\,\int_{S_{\bot}}\textrm{Res}_C[\omega_n^D] = (2\pi i)^{-\lceil c/2\rceil}\,\int_{\delta_CS_{\bot}}\omega_n^D\,,
\eeq
where $\delta_C \equiv \delta_{S_{1}\ldots S_{c}}$ denotes the iterated Leray coboundary. We then see that we can identify the integration contour $\Gamma_C$ with $\delta_CS_{\bot}$,
\beq\label{eq:final_cut_def}
\cC_CI_n =  (2\pi i)^{-\lceil c/2\rceil}\,\int_{\Gamma_C}\omega_n^D\,.
\eeq
To summarize, we have identified an integration contour that `encircles' all the poles of a given subset $C$ of propagators, making precise the (sometimes rather vague) definition of cuts in the literature. This contour is determined by the action of the iterated Leray coboundary on the intersection $S_C$ of the singular surfaces. The latter has the topology of a sphere. In the mathematical literature, $S_C$ and $\Gamma_C$ are sometimes referred to as the \emph{vanishing sphere} and the \emph{vanishing cycle} respectively.  This nomenclature will become clear in the next section.

\subsection{Polytope geometry and the Landau conditions}
\label{sec:polytopes}
\begin{figure}
\begin{centering}
\includegraphics[width=9cm]{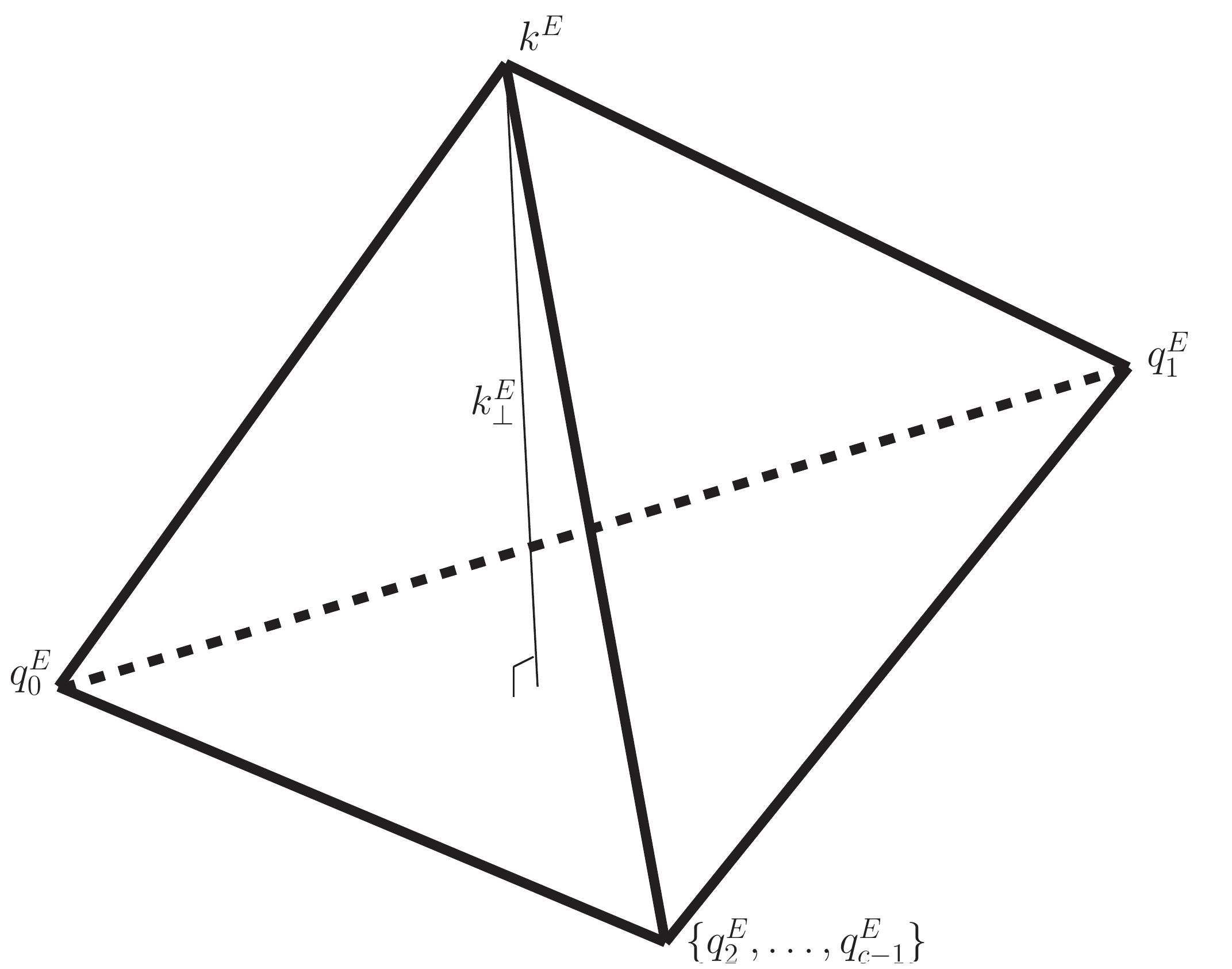}
\caption{The simplex $\mathcal{K}_{C}$ whose base is the simplex $\mathcal{Q}_{C}$, with the transverse component $k^E_\bot$ of the loop momentum. }
\label{fig:simplicesGen}
\end{centering}
\end{figure}

In this section we present a geometric interpretation of the final formula~\eqref{eq:cutdef-det} for one-loop cut integrals in terms of the geometry of the polytopes determined by the external momenta together with the loop momentum. Our polytope picture reproduces a similar geometric picture in the works of Cutkosky in the case of the three-point function~\cite{Cutkosky:1960sp}.
Let $\mathcal{Q}_{C}$ denote the $(c-1)$-simplex spanned by the edges $\{q^E_i-q^E_\ast:i\in C\setminus\{\ast\}\}$, arranged as vectors emanating from the common point $q^E_{\ast}$, $\ast\in C$.\footnote{At this point, we do not yet assume that the propagators in $C$ are cut.} Similarly, $\mathcal{K}_{C}$ denotes the $c$-simplex spanned by the edges $\{k^E-q^E_\ast,q^E_i-q^E_\ast:i\in C\setminus\{\ast\}\}$ (see \refF{fig:simplicesGen}).
Notice that the vertices of these simplices correspond directly to the dual momentum variables. For definiteness, we assume without loss of generality that $C=[c]$ and $q^E_\ast = q^E_1$.

By construction, the polytope $\mathcal{Q}_{C}$ lives in the linear subspace $\cE_C$, and it is a face of the simplex $\mathcal{K}_{C}$. The altitude of the polytope $\mathcal{K}_C$ above the face $\mathcal{Q}_{C}$ is given by $|k_{\bot}|=|k_\bot^E|$. Our goal is to describe the geometrical properties of these two polytopes. We define the following matrices whose columns are formed by our momentum vectors,
\bea
Q_{C} &=& \left(
q^E_2-q^E_1 \qquad q^E_3-q^E_1\quad
\cdots \quad
q^E_{c}-q^E_1
\right)\,, \\
K_{C} &=& \left(
k^E- q^E_1 \qquad q^E_2-q^E_1\quad
\cdots \quad
q^E_{c}-q^E_1
\right), \qquad K_{\emptyset} = \left( k^E \right)\,.
 \eea
Since it is preferable to work with Lorentz invariant expressions, we construct Gram determinants, and we can then write the volumes of the simplices in terms of  their square roots. In particular, the Gram determinants that we are going to consider are 
\bea
\Gram_{C} &=& \det Q_{C}^T Q_{C}, \qquad \Gram_{\emptyset} =  1\,, \\
H_{C} &=& \det K_{C}^T K_{C}\,.
\eea
It is easy to check that $\Gram_{C}$ and $H_{C}$ agree with the definitions of the Gram determinants in \refE{eq:gramHsubsetC} and \refE{eq:gramGsubsetC}.
The volumes of the simplices can then be written as
\bea
\vol \mathcal{Q}_{C} &=& \frac{1}{(c-1)!}\abs{\Gram_{C}}^{1/2}\,, \\
\vol \mathcal{K}_{C} &=& \frac{1}{c!}\abs{H_{C}}^{1/2}\,.
\eea
We now use the fact that the volume of an $N$-simplex can be computed in two equivalent ways:  first, by taking the determinant of the matrix whose columns are the edge vectors emanating from a common vertex, divided by $N!$, or second, by multiplying the volume of one of its $(N-1)$-dimensional faces by the altitude above that face, divided by $N$.
We conclude that the altitude of the polytope $\mathcal{K}_{C}$ above the face $\mathcal{Q}_{C}$ is given by
\beq
|k_{\bot}| = \left|\frac{H_{C}}{\Gram_{C}}\right|^{1/2}\,.
\eeq

So far, all the considerations are generic and apply independently of any propagators being cut. In the special case where the propagators from the set $C$ are cut, we know that $|Y_C|=|[H_C]_C|$, and we see that the height of $\mathcal{K}_{C}$ is fixed by the on-shell conditions in terms of the modified Cayley and Gram determinants,
\beq\label{eq:radius}
\left[|k_{\bot}|\right]_C = \left|\frac{Y_C}{\Gram_{C}}\right|^{1/2}\,.
\eeq
In other words, we see that the radius of the vanishing sphere $S_{\bot} = S_C$ is fixed by eq.~\eqref{eq:radius}.

In order to understand the geometric meaning of eq.~\eqref{eq:radius} and the nomenclature for the vanishing spheres $S_C$ and cycles $\Gamma_C$, it is useful to understand the connection between cut integrals and the Landau conditions~\cite{Landau:1959fi}, a set of necessary conditions on the external kinematics for a pinch singularity to occur.
If we work in Euclidean kinematics, then the Landau conditions for a one-loop integral  take the form
\bea
\alpha_i \left[(k^E-q^E_i)^2+m_i^2\right] = 0, \quad \forall i\,.
\label{eq:landau1}
\eea
and
\bea
\sum_{i=1}^{n} \alpha_i(k^E-q^E_i) = 0\,.
\label{eq:landau2}
\eea
The equations of the first Landau condition factorize:  for each $i$, either $\alpha_i=0$ or else the propagator is on shell.  Stated differently, if $C$ is the set of cut propagators, the first Landau condition can be satisfied by setting $\alpha_i=0$ for all $i \notin C$.\footnote{Landau originally insisted that all $\alpha_i \neq 0$; otherwise he considered it to be a singularity of a pinched graph rather than the original graph. The interest of configurations with some $\alpha_i=0$ was soon realized. In particular, the case where all but two $\alpha_i=0$ corresponds to discontinuities on physical channels~\cite{tHooft:1973pz,Veltman:1994wz,Cutkosky:1960sp}.}
After imposing the first Landau condition, the second Landau condition can be restated as
\bea
\sum_{i \in C} \alpha_i(k^E-q^E_i) = 0\,.
\label{eq:landau2cutset}
\eea
In order to characterize the solution space of \refE{eq:landau2cutset}, we can contract the equation with each momentum propagator $(k^E-q^E_j)$ for $j \in C$, giving the matrix equation
\beq\label{eq:system}
\left(\begin{array}{ccc}
(k^E-q^E_{1})\cdot (k^E-q^E_{1}) &  \ldots & (k^E-q^E_{1})\cdot (k^E-q^E_{c})\\
\vdots &\ddots & \vdots \\
(k^E-q^E_{c})\cdot (k^E-q^E_{1}) &  \ldots & (k^E-q^E_{c})\cdot (k^E-q^E_{c})
\end{array}\right) \left(\begin{array}{c}\alpha_{1}\\\vdots\\\alpha_{c}\end{array}\right) = 0\,.
\eeq
This linear system has a nontrivial solution only if its determinant vanishes. This determinant is the same as the Gram determinant $H_C$ defined in eq.~\eqref{eq:gramHsubsetC}, except that the momenta are Euclidean. We have seen that upon putting the propagators on-shell, $H_C$ becomes equal to the modified Cayley determinant $Y_C$. Equation~\eqref{eq:system} thus has a nontrivial solution if and only if $Y_C=0$. These solutions to the Landau conditions, corresponding to kinematic configurations where the modified Cayley determinant $Y_C$ vanishes, are called \emph{singularities of the first type}.\footnote{By the argument above, singularities of the first type comprise all solutions to the Landau conditions for finite values of loop momentum. Later we will also encounter singularities of the \emph{second type}. These are a separate class of solutions obtained at infinite loop-momentum, which are more conveniently analyzed in a different representation of Feynman integrals \cite{SMatrix}.}

There is a nice geometric interpretation of the Landau conditions at one loop. The on-shell conditions force the volume of the polytope $\mathcal{K}_C$ to be proportional to the (absolute value of the) modified Cayley determinant $Y_C$. Hence, we see that the volume of $\mathcal{K}_C$ vanishes at points where both Landau conditions are satisfied. 
Moreover, we know from eq.~\eqref{eq:radius} that the radius of the sphere $S_C$ is proportional to the square root of $Y_C$, and so we see that the radius of $S_C$ vanishes at the position of the pinch singularity. This is the reason for the names \emph{vanishing} sphere and cycle for $S_C$ and $\Gamma_C=\delta_CS_C$.

% !TEX root = main-cuts.tex

\section{Explicit results for some cut integrals}\label{sec:explicitRes}
In order to make the definitions of the previous section more concrete, we now present some explicit results for cut integrals. We give a concrete parametrization for the loop momentum with which the remaining integration in eq.~\eqref{eq:cutdef-det} can be carried out.  We identify conditions under which cuts can vanish identically and discuss their geometric interpretation. We then present some explicit results for maximal and next-to-maximal cuts.

\subsection{Evaluation of the residues}
\label{sec:parametrization}

We first introduce a parametrization of the loop momentum $k$ such that the residues factorize and can be computed sequentially. To align the indices of the $n$ propagators with those of the components of the momenta, we relabel the propagators from 0 to $n-1$ and using the $S_n$-symmetry of one-loop integrals, we assume without loss of generality that the set of cut propagators is $C=\{n-1,0,1,\ldots,c-2\}$, with all indices understood mod $n$. In this section, $C$ will always denote this set. We use momentum conservation to set $q^E_{n-1}=\bold{0}_D$.

Throughout this section, we work with the Euclidean momenta introduced in eq.~\eqref{eq:pre-euc-param} and in a frame where
\beq\label{eq:euc-param}
q_j^E=\left(q_{j0}^E,\ldots,q_{jj}^E,\mathbf{0}_{D-j-1}\right)\,, \qquad q_{jj}^E>0\,.
\eeq
We can parametrize the vector $k^E$ as
\begin{equation}
k^E=\rr\left(\cos\theta_0,\cos\theta_1\sin\theta_0,\ldots,\cos\theta_{n-2}\,\prod_{j=0}^{n-3}\sin\theta_j,\mathbf{1}_{D-n+1}\,\prod_{j=0}^{n-2}\sin\theta_j\right)\,,
\label{eq:angle-param}
\end{equation}
where $\mathbf{1}_{D-n+1}$ is a unit vector in $(D-n+1)$ dimensions, $r \geq 0$, and $0 \leq \theta_j < \pi$.
The integration in the remaining $(D-n+1)$ dimensions is trivial, and we obtain
the following integration measure,
\begin{equation}
\int d^Dk=i\int d^Dk^E=\frac{i\pi^{\frac{D-n+1}{2}}}{\Gamma\left(\frac{D-n+1}{2}\right)}\int d\rr^2 \left(\rr^2\right)^{\frac{D-2}{2}}\,\prod_{j=0}^{n-2}\int_0^\pi d\theta_j \sin^{D-2-j}\theta_j\,.
\end{equation}

Next,
for each angle $\theta_j$, we change variables to $t_j=(\cos\theta_j+1)/2$.
The propagators can be written as
\beq\bsp
(k^E-q^E_j)^2+m_j^2 =-A_j(\rr, t_0, t_1, \ldots t_{j-1})+t_j\, B_j(\rr, t_0, t_1, \ldots t_{j-1}) \, .
\label{eq:defAB}
\esp\eeq
We then obtain the following expressions for $A_j$ and $B_j$,
\begin{align}
A_j&=2\rr\left[\sum_{\alpha=0}^{j-1}q^E_{j\alpha}(2t_{\alpha}-1)\left(\prod_{\gamma=0}^{\alpha-1}2\sqrt{t_\gamma(1-t_\gamma)}\right)
-2^{j}q^E_{jj}\left(\prod_{\gamma=0}^{j-1}\sqrt{t_\gamma(1-t_\gamma)}\right)\right]\nonumber\\
&-m_j^2-(q^E_j)^2-\rr^2,\nonumber\\
{B}_j&=-2^{j+2}\rr q^E_{jj}\,\prod_{\beta=0}^{j-1}\sqrt{t_\beta(1-t_\beta)}\,.
\label{abeucliddefinition}
\end{align}
Equations~\eqref{eq:defAB} and~\eqref{abeucliddefinition} make manifest the main property of our parametrization that allows us to evaluate all the residues sequentially.
We see from \refE{eq:defAB} that the propagators have only simple poles in the variables $t_j$. Indeed, from \refE{abeucliddefinition} we see that
$A_j$ and $B_j$ only depend on the $t_\alpha$ with $\alpha<j$, and so the position of the poles of the cut propagators can easily be determined in terms of the variables $t_j$. It is then easy to see that in this parametrization the residues factorize, and so we can evaluate the residues sequentially at the simple poles in the $t_j$.
It will be useful to introduce the following notation for the positions of the poles, corresponding to the values of the $t_j$ where \refE{eq:defAB} vanishes,
\begin{align}\label{eq:t-poles}
T_{j}(r,t_0,\ldots,t_{j-1})=\frac{A_j(r,t_0,\ldots,t_{j-1})}{B_j(r,t_0,\ldots,t_{j-1})}\,.
\end{align}
In terms of the functions $T_{j}$, the propagators in \refE{eq:defAB} can be written as
\begin{equation}\label{eq:tp_def}
(k^E-q^E_j)^2+m_j^2=B_j(r,t_0,\ldots,t_{j-1})\,\left[t_j-T_{j}(r,t_0,\ldots,t_{j-1})\right]\,.
\end{equation}
Equation~\eqref{eq:tp_def} shows that in our
parametrization each propagator $(k^E-q^E_j)^2+m_j^2$ is naturally associated with a variable $t_j$ in which this propagator has a simple pole. The only exception is the
propagator $(n-1)$, which is associated to the radial coordinate $\rr$,
\begin{equation}
(k^E)^2+m_{n-1}^2=\rr^2+m_{n-1}^2\,.
\end{equation}

Before turning to the evaluation of the residues, it is instructive to see how the uncut integral looks in this parametrization. Since the integration contour is real and 
\beq\label{eq:sin_to_t}
\sin \theta_j = 2\sqrt{t_j(1-t_j)}\,,
\eeq 
all the $t_j$ must vary in the range $[0,1]$, and  the uncut integral can be written as
\begin{align}\label{eq:uncut_integral}
I_n=&(-1)^n\frac{2^{\sum_{j=0}^{n-2}(D-2-j)}e^{\gamma_{E}\epsilon}}{\pi^{\frac{n-1}{2}}\Gamma\left(\frac{D-n+1}{2}\right)}\int_0^\infty d\rr^2\frac{\left(\rr^2\right)^{\frac{D-2}{2}}}{\rr^2+m_{n-1}^2}
\prod_{j=0}^{n-2}\int_0^1 dt_j \frac{\left[t_j(1-t_j)\right]^{\frac{D-3-j}{2}}}{B_j\left( t_j-T_{j}\right)}\, ,
\end{align}
where we have dropped the dependence of the functions $B_j$ and $T_j$ on their arguments.

Let us now consider the cut integrals $\cC_CI_n$. We have already seen that $B_j$ and $T_{j}$ only depend on the radial coordinate $\rr$ and the $t_{\alpha}$ with $\alpha<j$. We can then easily evaluate the residues by starting from the integrand of the uncut integral, \refE{eq:uncut_integral}, and taking the residues at the poles of the cut propagators. In our parametrization, the position of the poles are defined iteratively by
\beq\label{eq:pole_position}
r^2=-m^2_{n-1} {\rm~~and~~} t_j = t_{j,p} \equiv T_j\left(\sqrt{-m_{n-1}^2},t_{0,p},\ldots,t_{j-1,p}\right)\,.
\eeq

The cut integral $\cC_CI_n$ is computed by sequentially taking the residues at each pole,
\begin{align}\label{cutdefeuclid}
\cutRes_{C}I_n&=
\frac{(2\pi i)^{\left\lfloor{c}/{2}\right\rfloor}\, {\rm e}^{\gamma_{E}\epsilon}}{\pi^{\frac{n-1}{2}}\Gamma\left(\frac{D-n+1}{2}\right)}
2^{\sum_{j=0}^{c-2}(D-2-j)}\,\, \times
\\&\nonumber
\res_{\rr^2=-m_{n-1}^2}\!\!\!\left[\res_{t_0=t_{0,p}}\!\!\!\left[\ldots
\res_{t_{m-2}=t_{m-2,p}}\!\!\!\left[\frac{\left(\rr^2\right)^{\frac{D-2}{2}}}{\rr^2+m_{n-1}^2}\prod_{j=0}^{c-2} \frac{\left[t_j(1-t_j)\right]^{\frac{D-3-j}{2}}}{B_j\left( t_j-T_{j}\right)}\,\,\, {f}^n_c\right] \right]\right]\,,
\end{align}
where ${f}^n_c$ collects the remaining $(n-c)$ integrations,
\begin{align}
\label{eq:fdefinition}
\begin{split}
{f}^n_c\left(\rr,t_0,\ldots,t_{c-2}\right)
&\equiv
2^{\sum_{j=c-1}^{n-2}(D-2-j)}\,\prod_{j=c-1}^{n-2}\int_0^1 dt_j \frac{\left[t_j(1-t_j)\right]^{\frac{D-3-j}{2}}}{B_j \left( t_j-T_{j}\right)}\,,
\end{split}
\end{align}
corresponding to the unconstrained variables $t_j$ with  $c-1\leq j\leq n-2$ that vary in the range $[0,1]$, just as in the case of the uncut integral in \refE{eq:uncut_integral}. We have suppressed the factor of $(-1)^n$ that appears in \refE{eq:uncut_integral} because we do not keep track of the overall sign of cut integrals, as discussed below \refE{eq:residue-def}. We will continue to suppress other powers of $-1$ in the equations that follow.

The residues in~\refE{cutdefeuclid} involve only simple poles, and so they can all be easily evaluated sequentially by eliminating 
the relevant denominators and substituting the values of $t_j$ at the poles in the remaining expression, see \refE{eq:pole_position}. We find that
\beq\bsp
\label{eq:simplepolesonly}
\cutRes_{C}I_n =
  (2\pi i)^{\left\lfloor{c}/{2}\right\rfloor}\frac{2^{\sum_{j=0}^{c-2}(D-2-j)}e^{\gamma_{E}\epsilon}}{\pi^{\frac{n-1}{2}}\Gamma\left(\frac{D-n+1}{2}\right)}\left(-m_{n-1}^2\right)^{\frac{D-2}{2}}\,[{f}^n_c]_C\,\prod_{j=0}^{c-2} \frac{\left[t_{j,p}(1-t_{j,p})\right]^{\frac{D-3-j}{2}}}{[B_{j}]_C}\,.
\esp\eeq
The function $f_c^n$ has been normalized such that in the case where all propagators are cut, $c=n$, we have $[f_n^n]_{C}=1$. Our goal is to identify $[{f}^n_c]_C$ with the integration over the vanishing sphere $S_C$ in eq.~\eqref{eq:cutdef-det},
\beq\bsp\label{eq:f_n_c}
[{f}^n_c]_C &\,= 2^{\sum_{j=c-1}^{n-2}(D-2-j)}\,\prod_{j=c-1}^{n-2}\int_0^1 dt_j \frac{\left[t_j(1-t_j)\right]^{\frac{D-3-j}{2}}}{[B_j]_C \left( t_j-[T_{j}]_C\right)}\\
&\,=\frac{\Gamma\left(\frac{D-n+1}{2}\right)}{2\pi^\frac{D-n+1}{2}} \int_{S_\bot} d\Omega^E_{D-c+1} \,\prod_{j\notin C} \left[\frac{1}{(k^E-q_j^E)^2+m_j^2}\right]_C\,.
\esp\eeq
In order to show that this is true, we need to show that the factors multiplying $[{f}^n_c]_C$ in eq.~\eqref{eq:simplepolesonly} combine to give the modified Cayley and Gram determinants in eq.~\eqref{eq:cutdef-det}.

\begin{figure}
\includegraphics[width=9cm]{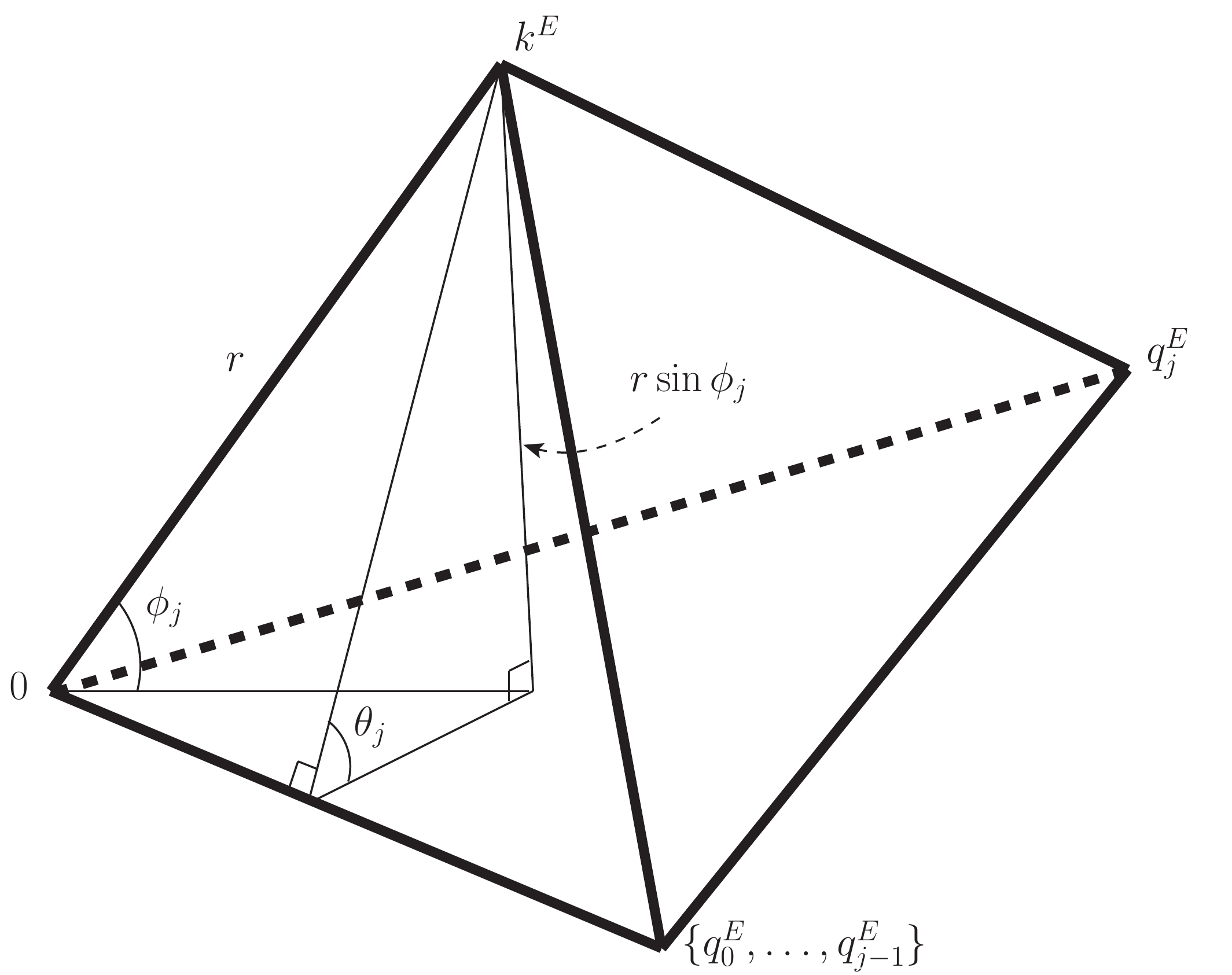}
\includegraphics[width=6cm]{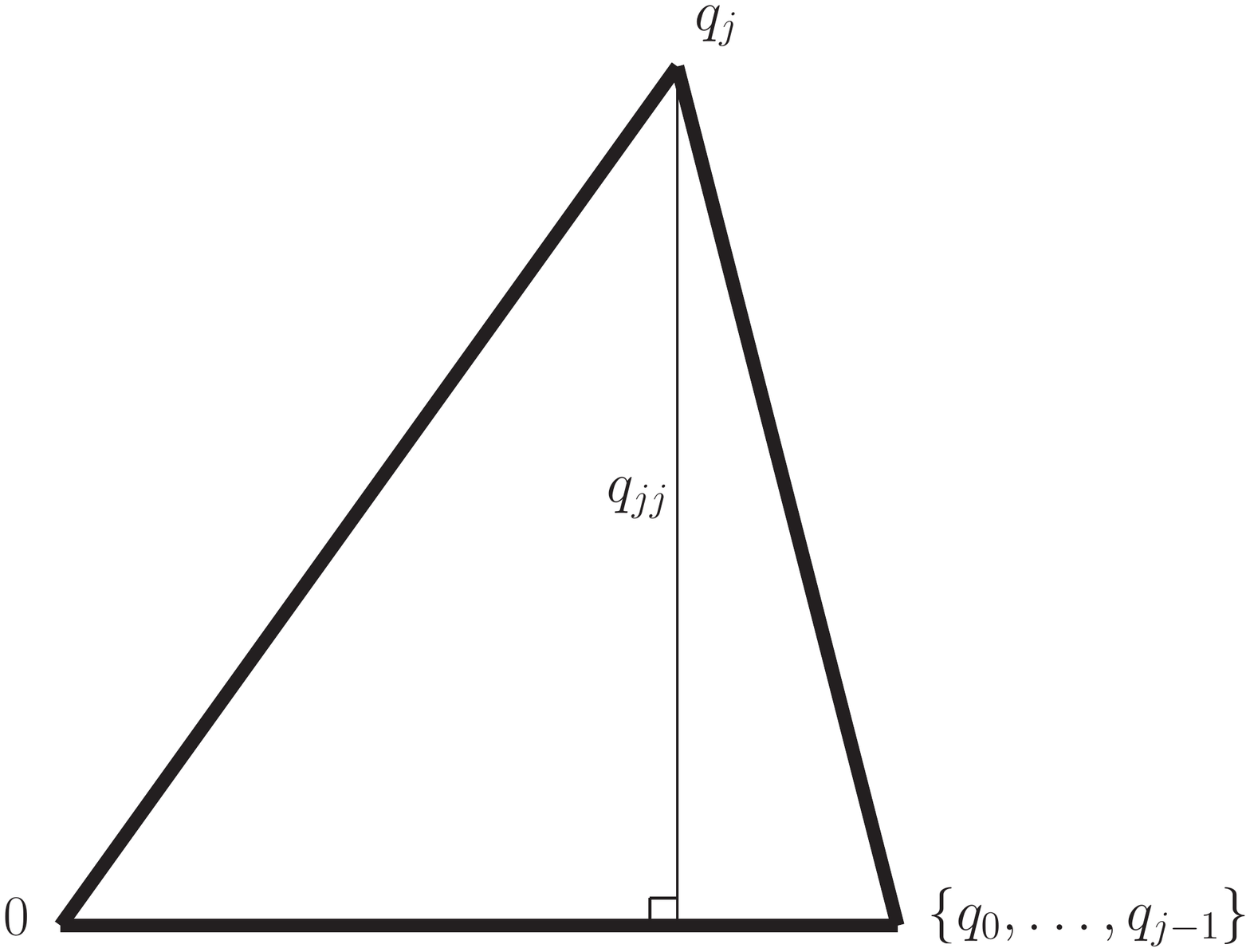}
\caption{The simplex $\mathcal{K}_{[k^E,0,1,2,\ldots,j]}$ and its base simplex $\mathcal{Q}_{[0,1,2,\ldots,j]}$, whose respective altitudes are $r\sin\phi_j$ and $q^E_{jj}$. }
\label{fig:simplices}
\end{figure}

We consider the polytope picture of cut integrals introduced in Section~\ref{sec:polytopes}, adapted to our choice of frame. We use the notation of Section~\ref{sec:polytopes}, set $q^E_{\ast}=q^E_{n-1}=0$, and keep $C=\{n-1,0,\ldots,c-2\}$ as above. Our goal is to express some of the components of the momenta in the frame defined in eqs.~\eqref{eq:euc-param} and \eqref{eq:angle-param} in terms of invariants.
% Because some components are complex (e.g., the variable $r$, as in  \refE{eq:pole_position}), the polytope picture only describes their magnitude, and we will reintroduce factors of $i$ at the end of our discussion.
Since the parametrization in eqs.~\eqref{eq:euc-param} and \eqref{eq:angle-param} relies on a specific ordering of the momenta $q^E_i$, we denote by $\mathcal{Q}_{[0,1,2,\ldots,j]}$ and $\mathcal{K}_{[k^E,0,1,2,\ldots,j]}$ the simplices with edges $\{q^E_0,\ldots,q^E_j\}$ and $\{k^E, q^E_0,\ldots,q^E_j\}$ respectively, with the edges given in this order. These simplices are represented in \refF{fig:simplices}.  

We observe that with the parametrization given in \refE{eq:euc-param} and \refE{eq:angle-param}, each angle $\theta_j$ can be interpreted as the dihedral angle between the faces $\mathcal{K}_{[k^E,0,1,2,\ldots,j-1]}$ and $\mathcal{Q}_{[0,1,2,\ldots,j]}$ in the simplex $\mathcal{K}_{[k^E,0,1,2,\ldots,j]}$.  If we further let $\phi_j$ be the angle between the edge $k^E$ and the face $\mathcal{Q}_{[0,1,2,\ldots,j]}$, then
\bea
\sin \phi_j = \prod_{k=0}^j \sin \theta_k.
\eea

Now we examine the equivalent formulas for computing volumes.  Consider the simplex $\mathcal{K}_{[k^E,0,1,2,\ldots,j]}$ and its face  $\mathcal{Q}_{[0,1,2,\ldots,j]}$.  The altitude of the simplex above this face is the distance from the endpoint of $k^E$ to $\mathcal{Q}_{[0,1,2,\ldots,j]}$, which is $\rr \sin\phi_j$.  Similarly, we consider the simplex $\mathcal{Q}_{[0,1,2,\ldots,j]}$ and its face $\mathcal{Q}_{[0,1,2,\ldots,j-1]}$.  The altitude of the simplex above this face is the distance from the endpoint of $q^E_j$ to $\mathcal{Q}_{[0,1,2,\ldots,j-1]}$, which is simply the (modulus of the) $j$-th component $q^E_{jj}$ of the vector $q^E_j$.  We obtain the following two relations,
\beq
\rr \sin \phi_j =  \frac{Y_{j+2}^{1/2}}{\Gram_{j+2}^{1/2}}\,,\qquad
q^E_{jj} = \frac{\Gram_{j+2}^{1/2}}{\Gram_{j+1}^{1/2}}\,,
\label{eq:altitudeQ}
\eeq
where we use the shorthand $Y_{j+2}\equiv Y_{\{n-1,0,\ldots,j\}}$, and similarly for $\Gram_{j+2}$. Note that $\rr \sin \phi_j$ is just the altitude $|k_\bot|$ defined in Section~\ref{sec:polytopes}.

In order to convert the expressions in
\refE{eq:simplepolesonly} to determinants, we observe
that $t_{a,p} (1-t_{a,p}) = \sin^2 \theta_j/4=\sin^2 \phi_j /(4 \sin^2 \phi_{j-1})$, and that $[{B}_{j}]_C =  -4\rr\sin\phi_{j-1} q^E_{jj}.$  As a consequence of the relations above, we can write
\beq\bsp\label{eq:variablesAsDets}
 \rr &= \,Y_{1}^{1/2}\,, \\
 t_{j,p} (1-t_{j,p})
 &= \frac{Y_{j+2} \Gram_{j+1}}{4 \Gram_{j+2} Y_{j+1}}\,, \\
 [{B}_{j}]_C
 &= -\,\frac{4 \Gram_{j+2}^{1/2} Y_{j+1}^{1/2}}{ \Gram_{j+1}}\, .
\esp\eeq
%where we have reintroduced a factor of $i$ in all terms proportional to $r$.
 It is now straightforward to derive the following result. 
 \bea\label{eq:simplepolesonly2}
\cutRes_{C}I_n&=&  (2\pi i)^{\left\lfloor {c}/{2}\right\rfloor}\frac{2^{1-c}e^{\gamma_{E}\epsilon}}{\pi^{\frac{n-1}{2}}\Gamma\left(\frac{D-n+1}{2}\right)}  
\frac{1}{\sqrt{\mu^c{Y_{C}}}}
\left(\mu\frac{Y_{C} }{\Gram_{C}} \right)^{\frac{D-c}{2}}
[{f}^n_c]_C \,.
\eea
Hence, comparing eq.~\eqref{eq:simplepolesonly2} to eq.~\eqref{eq:cutdef-det}, we conclude that eq.~\eqref{eq:f_n_c} is proven (we recall that eq.~\eqref{eq:f_n_c} only holds in Euclidean kinematics).

\subsection{Vanishing cuts}
\label{sec:vanishing}

We will now state several conditions sufficient for the vanishing of  one-loop cut integrals in dimensional regularization in generic kinematics, by which we mean that nonzero invariants are distinct.\footnote{If we were to consider degenerate kinematics where some internal and external masses become equal, we could have configurations with soft divergences. We would then find other conditions, beyond the ones listed below, under which two- or three-propagator cuts vanish.} We define $q_{ij}=q_i-q_j$, and we introduce the following shorthand for set-valued indices: if $i,\ldots,j$ are integers and $C$ is a set of integers, then  $\cC_{i\ldots j}I_n$ denotes $\cC_{\{i\ldots j\}}I_n$, and $\cC_{Ci}I_n$ denotes $\cC_{C\cup\{i\}}I_n$. Similar notation is used for the modified Cayley and Gram determinants.

We have identified three classes of vanishing cut integrals. In these cases, at most three propagators are cut:

\begin{enumerate}
\item Single-cut integrals vanish if the cut propagator is massless:
\beq
\cC_{i}I_n = 0\,, \quad \textrm{ if } m_i^2=0\,.
\eeq
\item A double-cut integral vanishes if the momentum flowing through the cut is lightlike:
\beq
\cC_{ij}I_n = 0\,, \quad \textrm{ if } q_{ij}^2=0\,.
\eeq
\item A triple-cut integral vanishes if the cut isolates a three-point vertex where three lightlike lines meet:
\beq
\cC_{ijk}I_n = 0\,, \quad \textrm{ if } q_{ij}^2=m_i^2=m_j^2=0 \,.
\eeq
\end{enumerate}

Let us discuss in turn the proofs of these claims, and let us start by showing that the single cut of a propagator of mass $m_i^2$ 
vanishes in the limit $m_i^2 \to 0$.
We have $Y_i = m_i^2$ and $\Gram_i=1$, and, using eqs.~\eqref{eq:cutdef-det} and \eqref{eq:f_n_c}, the resulting cut integral is
\begin{equation}\label{eq:vanishingOneCut}
	\cutRes_i I_n=\frac{e^{\gamma_E\eps}}{\pi^{\frac{n-1}{2}}\Gamma\left(\frac{D-n+1}{2}\right)}
	\left(-m_i^2\right)^{\frac{D-2}{2}}[f_1^n]_i\,.
\end{equation}
This integral will vanish in dimensional regularization, unless $[f_1^n]_i$ behaves like $(m_i^2)^{(2-D)/2}$ in the limit $m_i^2\to0$. If this were the case, then $[f_1^n]_i$ would either be divergent or vanish in the limit. We thus compute $[f_1^n]_i$ for $m_i^2=0$ and check that it is finite (neither zero nor divergent). From \refE{eq:fdefinition} we get
\begin{equation}
	[f_1^n]_i=2^{\sum_{j=0}^{n-2}(D-2-j)}
	\,\prod_{j=0}^{n-2}\int_0^1 dt_j \frac{\left[t_j(1-t_j)\right]^{\frac{D-3-j}{2}}}{\left[B_j\right]_i t_j-\left[A_{j}\right]_i}\,.
\end{equation}
For $m_i^2=0$, $[\cdot]_i$ means the quantities inside the bracket should be evaluated at $r=0$. From \refE{abeucliddefinition}, we get
\begin{equation}
	\left[A_{j}\right]_i=A_{j}(r=0)=-m_j^2-(q_j^E)^2 {\rm~~and~~} \left[B_j\right]_i=B_{j}(r=0)=0\,.
\end{equation}
Hence, the integration in eq.~\eqref{eq:f_n_c} can be done in closed form, and we obtain
\begin{equation}
	[f_1^n]_i=\pi^{\frac{n-1}{2}}
	\frac{\Gamma\left(\frac{D-n+1}{2}\right)}{\Gamma\left(\frac{D}{2}\right)}
	\prod_{j=0}^{n-2}\left(m_j^2+(q_j^E)^2\right)^{-1}\,,
\end{equation}
which shows that $[f_1^n]_i$ is well behaved as $m_i^2\to0$. This completes our proof of the vanishing of the massless single cut.

Next, let us consider the conditions for a double cut to vanish. Consider the cut of two propagators of masses $m_i^2$ and $m_j^2$, the difference of whose momenta is $q_{ij}$.  We have $\Gram_{ij}= q_{ij}^2$ and $Y_{ij}= \lambda(m_i^2,m_j^2,q_{ij}^2)/4$, where $\lambda$ denotes the K\"all\'en function,
\beq\label{eq:kallen}
\lambda(a,b,c) = a^2 + b^2 + c^2 - 2(ab+ac+bc)\,.
\eeq
The resulting cut integral is
\beq\bsp
\cC_{ij}I_n&\,= \frac{i\,\pi^{\frac{3-n}{2}}\,e^{\gamma_E\eps}}{\Gamma\left(\frac{D-n+1}{2}\right)}
\left[\frac{\lambda(m_i^2,m_j^2,q_{ij}^2)}{4} \right]^{\frac{D-3}{2}}
\,\left(q_{ij}^2\right)^{\frac{2-D}{2}}\,[f^n_2]_{ij}\,.
\esp\eeq
By an argument similar to the previous case, we can see that this expression vanishes in dimensional regularization as $q_{ij}^2\to0$. Alternatively, one can see that this conclusion is correct from \refE{cutdefeuclid}: if $q_{ij}^2=0$, then the corresponding propagator
\begin{equation}
	(k^E-q^E_{ij})^2+m_j^2=B_0(r)t_0-A_{0}(r)
\end{equation}
is actually independent of $t_0$ because $B_0(r)$ is zero for $q_{ij}^2=0$. In other words, there is no residue associated with $t_0$, and so the cut integral vanishes.

\begin{figure}
\center
\includegraphics[width=5cm]{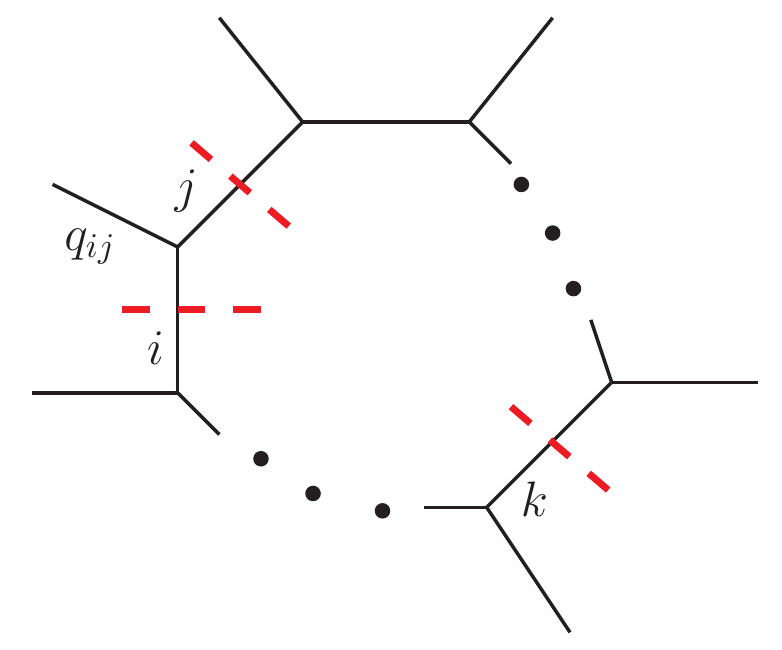}
\caption{Diagram representative of a triple cut.}
\label{fig:tripleCut}
\end{figure}

Finally, let us turn to the vanishing of the triple cut. Consider the case where three propagators are cut, as depicted in \refF{fig:tripleCut}, in a way that isolates a three-point vertex connecting three massless lines (in \refF{fig:tripleCut}, the vertex connecting the lines with labels $i$, $j$ and $q_{ij}$).  At one loop, the three massless lines must include two adjacent massless propagators, labeled by $i$ and $j$, and the external massless line incident to their common vertex carries the momentum $q_{ij}$, such that $q_{ij}^2=0$.  The choice of the third cut propagator is unimportant, and it is labeled by $k$. The corresponding modified Cayley matrix takes the form
\bea\label{eq:Y_triple_cut}
Y_{ijk}=
\det\left(
\begin{array}{ccc}
0 & 0 & \frac{m_k^2-q^2_{ik}}{2} \\
0 & 0 & \frac{m_k^2-q^2_{jk}}{2} \\
\frac{m_k^2-q^2_{ik}}{2} & \frac{m_k^2-q^2_{jk}}{2} & m_k^2
\end{array}
\right) = 0\,.
\eea
The vanishing of the determinant in eq.~\eqref{eq:Y_triple_cut} is due to a collinear singularity at the vertex where propagators $i$ and $j$ meet \cite{Ellis:2007qk}.
To conclude that the cut integral vanishes, we follow the same path as for one-propagator cuts and show that $[f^n_3]_{ijk}$ is regular in this limit (i.e., it neither vanishes nor diverges). To make the connection with the discussion in the previous section, we should relabel indices so that in \refF{fig:tripleCut} the propagators $i$, $j$ and $k$ become respectively the propagators with index 0, 1 and $n-1$. 
Then, the collinearity condition corresponds to $\sin\theta_1=0$, which through eq.~\eqref{eq:variablesAsDets} implies\footnote{We can check by explicit calculation that in this collinear configuration either $[B_1]_{ijk}\neq0$ and $[A_1]_{ijk}=0$, or $[A_1]_{ijk}=[B_1]_{ijk}$, depending on the details of the parametrization.} that $t_{1,p}(1-t_{1,p})=0$. We then have
\begin{equation}
	[B_2]_{ijk}=-2^4rq_{22}^E\sqrt{t_{0,p}(1-t_{0,p})}\sqrt{t_{1,p}(1-t_{1,p})}=0\,,
\end{equation}
which in turn implies that $[B_\alpha]_{ijk}=0$ for all $\alpha\geq 2$. The integration of the $t_\alpha$ in $[f^n_3]_{ijk}$ can thus be done in closed form, and we find
\begin{equation}
	[f^n_3]_{ijk}=\pi^\frac{n-3}{2}
	\frac{\Gamma\left(\frac{D-n+1}{2}\right)}{\Gamma\left(\frac{D-2}{2}\right)}
	\prod_{j=2}^{n-2}\left(-[A_\alpha]_{ijk}\right)^{-1}\,.
\end{equation}
The last expression is finite, so the limit $Y_{ijk}\to0$ of $[f^n_3]_{ijk}$ is well defined. This completes the proof of the vanishing of three-propagator cuts that isolate a massless three-point vertex.

The conditions listed above for one-propagator and three-propagator cuts illustrate that the vanishing of the vanishing sphere (cf.~\refE{eq:radius}) leads to a vanishing cut.  In this section, we have simply checked that this conclusion still holds in dimensional regularization when some scales are zero. On the other hand, the two-propagator cut vanishes when the corresponding Gram determinant does: this condition will be analyzed in  \refS{sec:secondType} in the context of second-type singularities.

We would like to emphasize that cuts such as the quadruple cut of a massless box do not fall into any of the above categories, and in fact this particular cut is nonzero.  All of its  three-propagator cuts vanish, for the reason given above. On the other hand, for the quadruple cut we have $Y_{[4]}=s^2t^2/16$ and $\Gram_{[4]}=-st(s+t)/4$, and thus we find
\begin{equation}
\label{eq:max-cut-box}
	\cutRes_{[4]}\tildeJ_4^{0-\textrm{mass}}=2e^{\gamma_E\epsilon}\frac{\Gamma(1-\epsilon)}{\Gamma(1-2\epsilon)}
	\frac{(s+t)^\eps}{(s\,t)^{1+\epsilon}}\,.
\end{equation}
More generally, we have checked explicitly that $Y_C$ and $\Gram_C$ do not vanish for $4\leq|C|\leq 8$ in generic kinematic configurations, even in the case where all propagators and external legs are massless. Based on this observation, we conjecture that cuts of four or more propagators never vanish in generic kinematic configurations.

\subsection{Explicit results for maximal cuts}
\label{sec:max_cut}
In this section we show that
the general formula~\eqref{eq:cutdef-det} takes a particularly simple form when all propagators are cut, the so-called \emph{maximal cut}. 
In this case, the remaining angular integration in eq.~\eqref{eq:cutdef-det} is trivial, because the integrand does not contain any additional propagators. 
We look specifically at the class of one-loop integrals in even dimensions nearly matching the number of propagators, as specified in \refE{eq:tildeJ_n}. In Section~\ref{sec:ibps} we will see that these results suffice to compute the maximal cuts of arbitrary one-loop integrals.

We label the propagators from 1 to $n$ and
distinguish the cases of $n$ even and $n$ odd. For $n$ even, we find
\beq\bsp\label{eq:MaxCutEven}
\cC_{[n]}&\tildeJ_{n}= 2^{1-2\eps-\frac{n}{2}}i^{\frac{n}{2}}\,\frac{e^{\gamma_E\eps}\Gamma(1-\eps)}{\Gamma(1-2\eps)}\,
\frac{1}{\sqrt{Y_{[n]}}}
\left(\frac{Y_{[n]}}{\Gram_{[n]}}\right)^{-\eps}\,,\qquad n\textrm{ even.}
\esp\eeq
For $n$ odd, we find,
\beq\bsp\label{eq:MaxCutOdd}
\cC_{[n]}&\tildeJ_{n}= 2^\frac{1-n}{2}i^{\frac{n-1}{2}}\,\frac{e^{\gamma_E\eps}}{\Gamma(1-\eps)}\frac{1}{\sqrt{\Gram_{[n]}}}\left(\frac{Y_{[n]}}{\Gram_{[n]}}\right)^{-\eps}\,,\qquad n\textrm{ odd.}
\esp\eeq

In the previous section we have shown that triple cuts that isolate a three-point vertex where three massless lines meet vanish in dimensional regularization. As a consequence, a triangle integral with two massless propagators such that the difference of momenta flowing through them is massless must have a vanishing maximal cut (they correspond to cases where $Y_{[3]}=0$ in \refE{eq:MaxCutOdd}). These triangle integrals are precisely those that are reducible to lower point integrals. Conversely, we have argued that cut integrals with four or more cut propagators never vanish in generic kinematics, and so we expect that integrals with four or more propagators cannot be reduced to lower point integrals.

\subsection{Explicit results for next-to-maximal cuts }\label{sec:nmaxCuts}
Cut integrals with all but one of the propagators cut, the so-called \emph{next-to-maximal cuts}, also admit a particularly simple closed expression. 
We first discuss a general integral $I_n$, and then restrict our analysis to the basis integrals $\tildeJ_n$. The discussion in this section is valid if the maximal cut does not vanish. 

We label the propagators from 1 to $n$ and assume without loss of generality that the set of cut propagators is $C=[n-1]$.
In the case of a next-to-maximal cut the remaining angular integral can be carried out in terms of Gauss's hypergeometric function,
\beq
{{}_2F_1}(a,b;c;z) = \frac{\Gamma(c)}{\Gamma(b)\Gamma(c-b)}\,\int_0^1 dt\,t^{b-1}\,(1-t)^{c-b-1}\,(1-zt)^{-a}\,.
\eeq
From eq.~\eqref{eq:f_n_c}, we find
\begin{equation}\label{eq:nmaxF}
	[f^n_{n-1}]_C=-\frac{2^{D-n}\Gamma^2\left(\frac{D-n+1}{2}\right)}{t_{p}\,[B]_C\,\Gamma(D-n+1)}
	\hypgeo{1}{\frac{D-n+1}{2}}{D-n+1}{\frac{1}{t_{p}}}\,.
\end{equation}
Using the relation~\cite{handbook}
\begin{equation}\label{eq:hypgeorel}
	\hypgeo{a}{b}{2b}{z}=(1-z)^{-\frac{a}{2}}
	\hypgeo{\frac{a}{2}}{b-\frac{a}{2}}{b+\frac{1}{2}}{-\frac{z^2}{4(1-z)}}\,,
\end{equation}
and the fact that \refE{eq:variablesAsDets} implies
\begin{equation}\label{eq:temp1320}
t_{p}(1-t_{p})=\frac{Y_{[n]}\Gram_{[n-1]}}{4\Gram_{[n]}Y_{[n-1]}}\,,
\end{equation}
we observe that the left-hand side of eq.~\eqref{eq:nmaxF} can be written entirely in terms of Gram and modified Cayley determinants:
\begin{equation}\label{eq:nmaxNotPractical}
	[f^n_{n-1}]_C\!=\!\frac{2^{D-n-1}\Gamma^2\left(\frac{D-n+1}{2}\right)}{\Gamma(D-n+1)}\sqrt{-\frac{\Gram_{[n-1]}}{Y_{[n]}}}\!
	\hypgeo{\frac{1}{2}}{\frac{D-n}{2}}{\frac{D-n+2}{2}}{\frac{\Gram_{[n]}Y_{[n-1]}}{\Gram_{[n-1]}Y_{[n]}}\!}\!.
\end{equation}
Inserting this result into \refE{eq:cutdef-det}, we easily find that the next-to-maximal cut of a diagram with $n$ propagators is 
\begin{align}\bsp\label{eq:NcutRes}
    \cutRes_{[n-1]}I_n=&\,(2\pi i)^{\lfloor\frac{n-1}{2}\rfloor}\frac{2^{D-2n+1}e^{\gamma_E\eps}\Gamma\left(\frac{D-n+1}{2}\right)}{\pi^{\frac{n-1}{2}}\Gamma(D-n+1)}
    \frac{1}{\sqrt{-Y_{[n]}}}\left(\frac{Y_{[n-1]}}{\Gram_{[n-1]}}\right)^{\frac{D-n}{2}}\\
	&\hypgeo{\frac{1}{2}}{\frac{D-n}{2}}{\frac{D-n+2}{2}}{\frac{\Gram_{[n]}Y_{[n-1]}}{\Gram_{[n-1]}Y_{[n]}}}.
\esp\end{align}

While the hypergeometric function in eq.~\eqref{eq:NcutRes} has a simple argument, its indices are in general half integers for $\eps\to0$, so it is not directly obvious that the Laurent expansion in dimensional regularization can be expressed in terms of polylogarithmic functions. We next present an alternative way to write the next-to-maximal cut of the basis integrals $\tildeJ_n$ where all the indices of the ${}_2F_1$ function are integers as $\eps\to0$, and so all the coefficients in the Laurent expansion are polylogarithmic. The price to pay is that the arguments of the polylogarithms involve square roots. Equation~\eqref{eq:temp1320} implies that
\begin{equation}
t_{p}=\frac{1}{2}\left(1\pm\sqrt{1-\eta}\right)\,,
\end{equation}
where we define
\begin{equation}
\eta\equiv\frac{Y_{[n]}\Gram_{[n-1]}}{\Gram_{[n]}Y_{[n-1]}}\,.
\end{equation}
For concreteness, we assume in the following that $0<\eta<1$ and choose the solution of \refE{eq:temp1320} with a plus sign. The results of other choices can be obtained by analytic continuation, and they would be equivalent modulo $i\pi$.

Let us separate the cases where $n$ is even or odd. 
If $n$ is odd, \refE{eq:nmaxF} immediately gives
\begin{equation}\bsp\label{eq:NMaxCutOdd}
    \cutRes_{[n-1]}\tildeJ_n&\,=-2^{\frac{3-n}{2}-2\eps}i^{\frac{n-1}{2}}\frac{e^{\gamma_E\eps}\Gamma(1-\eps)}{\Gamma(2-2\eps)}\left(\frac{{Y_{[n-1]}}}{{\Gram_{[n-1]}}}\right)^{-\epsilon}\frac{1}{\sqrt{\Gram_{[n]}}}\\
    &\,\times \frac{1}{1+\sqrt{1-\eta}}\hypgeo{1}{1-\epsilon}{2-2\epsilon}{\frac{2}{1+\sqrt{1-\eta}}},
    \qquad\qquad n \textrm{ odd}\,.
\esp\end{equation}
We see that the hypergeometric function has integer indices. Its Laurent expansion in $\eps$ can therefore be expressed in terms of polylogarithms to all orders~\cite{Moch:2001zr}. The first
order is:
\begin{equation}
	\!\!\cutRes_{[n-1]}\tildeJ_n\!=\!
	\frac{2^{\frac{1-n}{2}}i^{\frac{n-1}{2}}}{\sqrt{\Gram_{[n]}}}
	\ln\left(
	\frac{\sqrt{Y_{[n]}\Gram_{[n-1]}-\Gram_{[n]}Y_{[n-1]}}-\sqrt{-\Gram_{[n]}Y_{[n-1]}}}{\sqrt{Y_{[n]}\Gram_{[n-1]}-\Gram_{[n]}Y_{[n-1]}}+\sqrt{-\Gram_{[n]}Y_{[n-1]}}}
	\right)+\mathcal{O}(\eps)
\end{equation}

If $n$ is even, a convenient representation of the hypergeometric function is obtained by applying the following transformation~\cite{handbook} to~\refE{eq:nmaxF},
\begin{equation}
	\hypgeo{a}{b}{2b}{z}=\left(\frac{1+\sqrt{1-z}}{2}\right)^{-2a}
	\hypgeo{a}{a-b+\frac{1}{2}}{b+\frac{1}{2}}{\left(\frac{1-\sqrt{1-z}}{1+\sqrt{1-z}}\right)^2}\,.
\end{equation}
The next-to-maximal cut of $\tildeJ_n$ with $n$ even is then given by
\begin{equation}\bsp\label{eq:NMaxCutEven}
    \cutRes_{[n-1]}\tildeJ_n&\,=\,-2^{1-\frac{n}{2}}i^{\frac{n-2}{2}}\frac{e^{\gamma_E\eps}}{\Gamma(1-\eps)}\left(\frac{{Y_{[n-1]}}}{{\Gram_{[n-1]}}}\right)^{-\epsilon}
    \frac{\sqrt{\Gram_{[n-1]}}}{\sqrt{\Gram_{[n]}Y_{[n-1]}}}\frac{1}{\sqrt{1-\eta}+\sqrt{-\eta}}\\
    &\,\times \hypgeo{1}{1+\epsilon}{1-\epsilon}{\frac{\sqrt{1-\eta}-\sqrt{-\eta}}{\sqrt{1-\eta}+\sqrt{-\eta}}},
    \qquad\qquad n \textrm{ even}\,.
\esp\end{equation}
We again see that we can express the next-to-maximal cut in terms of hypergeometric functions with integer indices only. The first orders are
\begin{align}
		\cutRes_{[n-1]}\tildeJ_n=&
		-\frac{2^{-\frac{n}{2}}\,i^{\frac{n}{2}}}{\sqrt{Y_{[n]}}}
		\left(
		1+2\eps\ln\left(1+\sqrt{\frac{Y_{[n]}G_{[n-1]}-G_{[n]}Y_{[n-1]}}{Y_{[n]}G_{[n-1]}}}\right)+\right.\\
		&\left.\eps\ln\left(\frac{G_{[n-1]}}{4\,Y_{[n-1]}}\right)
		\right)+\mathcal{O}\left(\eps^2\right)\,.\nonumber
\end{align}

Let us conclude this section by highlighting a relation between the maximal and next-to-maximal cuts of a one-loop integral with an even number of propagators.
Expanding the explicit results for these cuts in eqs.~\eqref{eq:MaxCutEven} and \eqref{eq:NMaxCutEven} to leading order in the dimensional regulator, we find
\begin{equation}
\label{eq:Max-NMax-2_abs}
    \cC_{[n-1]}\tildeJ_{n} = -\half\cC_{[n]}\tildeJ_{n} + \ord(\eps)\,\,,
    \quad n \textrm{ even}\,.
\end{equation}
In principle, the sign in this relation is not determined, as our cuts are defined up to an overall sign as discussed below \refE{eq:cutdef-det}. Imposing the minus sign in \refE{eq:Max-NMax-2_abs} fixes the relative sign between maximal and next-to-maximal cuts of integrals with even number of propagators. In the next section, we will see this relation is a special case of a larger class of relations between cut integrals. Finally, we stress that \refE{eq:Max-NMax-2_abs} only holds if both sides of the equality are well defined at $\eps=0$.

% !TEX root = main-cuts.tex

\section{Compactification and singularities of the second type}
\label{sec:secondType}

In addition to pinch singularities, Feynman integrals may also exhibit singularities when the integration contour is pinched at infinity~\cite{SMatrix,SecondType1,SecondType2}. These so-called \emph{singularities of the second type}  are classified by the vanishing of the Gram determinant $\Gram_C$ (see Appendix~\ref{app:Landau} for a derivation) rather than the modified Cayley determinant,\footnote{Accordingly, singularities of the second type also have a nice interpretation in the context of the polytope geometry discussed in Section~\ref{sec:polytopes}. We have seen that singularities of the first type correspond to the degeneration of the polytope $\mathcal{K}_C$ due to the vanishing of the altitude above its face $\mathcal{Q}_C$. Singularities of the second type can be seen as the degenerations of $\mathcal{K}_C$ due to a vanishing of the volume of its face $\mathcal{Q}_C$.} and they are not directly related to the cut integrals as defined above.

However, the notion of cut integrals can indeed be extended to singularities of the second type. In this section, we will make this statement precise by performing a compactification and considering residues at infinity.  In the compactified picture, singularities of the first and second types can be treated on the same footing. Furthermore, for one-loop integrals, the so-called Decomposition Theorem~\cite{PhamCompact,Froissart,Teplitz} implies that cut integrals for singularities of the two  types are not independent, leading to various linear relations among cut integrals.

\subsection{Compactification of one-loop integrals}
\label{sec:compactification}

In previous sections, we have seen that to every singularity of the first type defined by $Y_C=0$ we can associate the cut integral $\cC_CI_n$. In order to extend these concepts to the singularities of the second type, the usual momentum representation of loop integrals is not the most convenient, because the pinch happens at infinite loop momentum. We therefore use a representation of one-loop integrals as integrals over a compact quadric in the complex projective space $\mathbb{CP}^{D+1}$~\cite{PhamCompact,SimmonsDuffin:2012uy,Caron-Huot:2014lda}, which we review in this subsection. Throughout this section we work in Euclidean kinematics, and we strictly follow the conventions of ref.~\cite{SimmonsDuffin:2012uy}. We write points $Z$ in $\mathbb{CP}^{D+1}$ in terms of homogeneous coordinates as 
\beq
Z = \left[\begin{array}{c}z^\mu\\ Z^-\\ Z^+\end{array}\right]\,,
\eeq
 and we identify $Z$ and $\alpha\,Z$, with $\alpha\in\mathbb{C}$, $\alpha\neq0$. We equip $\mathbb{CP}^{D+1}$ with the bilinear form
 \beq\label{eq:scalarProd}
 (Z_1Z_2) = z_1^\mu\,z_{2\mu} - \frac{1}{2}\,Z_1^+Z_2^- - \frac{1}{2}\,Z_1^-Z_2^+\,,
 \eeq
 where $z_1^\mu\,z_{2\mu}$ denotes the usual Euclidean scalar product.
 If we work in the coordinate patch $Z^+=1$, then to each propagator $D_i = (k^E-q^E_i)^2+m_i^2$ we associate  the point $X_i\in \mathbb{CP}^{D+1}$ defined by
\beq
X_i = \left[\begin{array}{c}(q^E_i)^\mu\\ (q_i^E)^2+m_i^2\\ 1\end{array}\right]\,,\qquad 1\le i\le n\,.
\eeq
Note that $(X_iX_i)=-m_i^2<0$ for positive values of the masses.
The one-loop integral $I_n^D$ can then be written as an integral over the $D$-form $\varpi_n^D$,
\beq\label{eq:compactified}
I_n^D = \int_\Sigma\varpi_n^D\,,\textrm{ with }\varpi_n^D =  \frac{(-1)^n\,e^{\gamma_E\eps}}{\pi^{D/2}}\frac{d^{D+2}Y\,\delta((YY))}{\textrm{Vol}(GL(1))}\frac{[-2(X_{\infty}Y)]^{n-D}}{[-2(X_1Y)]\ldots[-2(X_{n}Y)]}\,,
\eeq
where ${\Sigma}$ is the real quadric defined by $(YY)=0$ (take e.g.~$Y=[(k^E)^\mu,(k^E)^2,1]^T$), and we have  introduced the `point at infinity'
\beq\label{eq:infinityPoint}
X_{\infty}= \left[\begin{array}{c}0^\mu\\ 1\\ 0\end{array}\right]\,.
\eeq
The singular surfaces where the propagators go on shell are mapped to the hyperplanes $P_i$ in $\mathbb{CP}^{D+1}$ defined by $(X_iY)=0$, $i\in[n]$. There is an additional singular hyperplane $P_{\infty}$ defined by $(X_{\infty}Y)=0$. In the compactified picture, the solutions to the Landau conditions are classified by the vanishing of the Gram determinants $\det(X_iX_j)_{i,j\in C}$, where
now $C$ is a subset of $\{1,\ldots,n,\infty\}$.
These determinants are related to the original Gram and modified Cayley  determinants defined in \refE{eq:gramGsubsetC} and \refE{eq:mcayleyYsubsetC} by
\beq\label{eq:compact_Gram}
\det(X_iX_j)_{i,j\in C} = \left\{\begin{array}{ll}
(-1)^c\,Y_C\,,&\textrm{ if } \infty \notin C\,,\\
\frac{(-1)^{c-1}}{4}\,\Gram_{C\setminus \{\infty\}}\,,&\textrm{ if } \infty \in C\,.
\end{array}\right.
\eeq
After compactification, there is no distinction between the Landau singularities of the first and second type: they are treated on the same footing. The compactification of one-loop integrals thus provides the ideal framework to extend the notion of cut integrals to Landau singularities of the second type. However, there are two issues which we need to address in order to precisely define the cut integrals associated with these singularities:
\begin{enumerate}
\item We need to know the vanishing spheres and cycles in the compactified picture, because they provide the integration contours for cut integrals.
\item In dimensional regularization, the integrand $\varpi_n^D$ has a `branch point' on $P_{\infty}$ rather than a pole, as seen from the factor of $(X_\infty Y)^{n-D}$. Hence, we cannot apply the residue theorem~\eqref{eq:residue_thm} which was crucial in the derivation of eq.~\eqref{eq:final_cut_def}.
\end{enumerate}
Both of these issues have been solved in the mathematical literature, and we review these solutions in the next sections. 

\subsection{Homology groups associated to one-loop integrals}
\label{sec:homology}

Working in the compactified picture, we need to identify the vanishing spheres $\tilde{S}_C$ and vanishing cycles $\tilde{\Gamma}_{C}$ in order to understand one-loop cut integrals, according to the discussion of Section~\ref{sec:residues}. We use a tilde to distinguish the vanishing spheres and cycles in the compactified picture from those in the momentum space picture, and we denote the complex quadric $(YY)=0$ by $\overline{\Sigma}$, in contrast to the real quadric $\Sigma$ of the previous subsection.

 The vanishing spheres and cycles are most conveniently characterized by studying the homology groups of $\overline{\Sigma} - P_{\infty} - P^{[n]}$ where 
\beq
P^C \equiv \bigcup_{j\in C}P_J {\rm~~and~~}P_C \equiv \bigcap_{j\in C}P_J\,,\qquad C\subseteq [n]\cup\{\infty\}\,.
\eeq
Very loosely speaking, homology groups classify all the non-equivalent integration contours that we can define on a space. The homology groups relevant to one-loop integrals\footnote{The homology groups are specific to the compactification, and it is not clear if different compactifications may lead to different homology groups~\cite{Teplitz}.} have been studied in ref.~\cite{PhamCompact,Froissart,Teplitz}, where it was shown that they are one-dimensional and generated by all the cycles that wind around\footnote{We recall that we work in a complexified space, where for each point of a hypersurface we can define a complex plane transverse to the hypersurface $P_j$. We can then consider a loop around this point in the transverse space that encircle  the hyperplane $P_j$ without touching it.} $P_j$, with  $j\neq\infty$. The rest of this subsection consists of a short review of this result.

We identify $\mathbb{C}^{D+1}$ with the coordinate patch $Z^+=1$ and define $\overline{\Sigma}'\equiv \overline{\Sigma}\cap\mathbb{C}^{D+1}$. We will be interested in the surfaces $\overline{\Sigma}'\cap P_{C}$. Since $\overline{\Sigma}'$ is a quadric and $P_{C}$ is an intersection of hyperplanes, each $\overline{\Sigma}'\cap P_{C}$ is a projective $(D-c)$-dimensional quadric, with $c=|C|$, which is necessarily compact. Consider now the homology groups $H_{D-c}(\overline{\Sigma}'\cap P_{C})$: they are one-dimensional and generated by elements\footnote{Strictly speaking, the elements of the homology groups are equivalence classes of cycles, where two cycles are equivalent if they differ by a boundary.} of the form $\overline{\Sigma}'\cap P_{C}$. The real points of $\overline{\Sigma}'\cap P_{C}$ must correspond to the vanishing sphere $\tilde{S}_{C}$, because by definition the vanishing sphere associated to a subset of propagators going on shell is contained in the intersection $P_{C}$ of the corresponding singular surfaces. In other words, in the compactified picture the vanishing spheres $\tilde{S}_C$ are simply the intersection of the hyperplanes $P_j$, $j\in C$, and the quadric ${\Sigma}$. We can now construct the vanishing cycles $\tilde\Gamma_C$ associated with the vanishing spheres $\tilde{S}_{C}$: if $\delta_{C}$ is the iterated Leray coboundary associated to the singular surfaces $P_j$, $j\in C$, then we set $ \tilde{\Gamma}_{C}\equiv \delta_{C}\tilde{S}_{C} =\delta_{C}({\Sigma}\cap P_{C}) $  and $\tilde\Gamma_{\emptyset} \equiv \Sigma$. Note that the antisymmetry of the iterated Leray coboundary, eq.~\eqref{eq:Leray_antisymmetry}, implies that $\tilde{\Gamma}_{C}$ is only defined up to a sign coming from the ordering of the singular surfaces.

It can then be shown that
\beq\bsp\label{eq:homology}
H_D(\overline{\Sigma} - P_{\infty} - P^{[n]}) &\,=
H_D(\overline{\Sigma}') \oplus \bigoplus_{\emptyset\subset C\subseteq {[n]}}\delta_{C}H_{D-c}(\overline{\Sigma}'\cap P_{C})\,.
\esp\eeq
This result is known as the Decomposition Theorem~\cite{PhamCompact,Froissart,Teplitz}.
Since the right-hand side of eq.~\eqref{eq:homology} does not involve the hyperplane $P_\infty$, the structure of the homology group $H_D(\overline{\Sigma} - P_{\infty} - P^{[n]})$ associated to $I_n^D$ is determined entirely by the singular surfaces associated to the propagators of the one-loop integral. Thus  the homology group associated to $I_n^D$ is generated by the vanishing cycles $\tilde{\Gamma}_C$ (which in the momentum space picture correspond to the contours that encircle a specific set of propagator poles).

The singularities associated with the hyperplane $P_{\infty}$ give rise to additional vanishing cycles $\tilde\Gamma_{\infty C}$. It follows from \refE{eq:homology} that we must be able to write each $\tilde\Gamma_{\infty C}$ as a linear combination of the basis $\{\tilde{\Gamma}_C: C\subseteq [n]\}$. Explicitly, one finds~\cite{PhamInTeplitz},
\beq\bsp
\label{eq:hom_rel}
\tilde\Gamma_{\infty C} &\,= -2x_{c}\,\tilde\Gamma_{C}-\sum_{C\subset X\subseteq [n]} (-1)^{\lceil \abs{C}/2\rceil+\lceil\abs{X}/2\rceil}\,\tilde\Gamma_X\,,
\esp\eeq
where
\beq\label{eq:xn_def}
x_{c} = \left\{\begin{array}{ll}
1\,, &\textrm{ if } c \textrm{ odd}\,,\\
0\,, &\textrm{ otherwise}\,.
\end{array}\right.
\eeq
Note that for eq.~\eqref{eq:hom_rel} to hold, we need a specific choice for the signs coming from the ordering of the singular surfaces. From now on, unless stated otherwise, we assume that all signs are fixed in such a way that eq.~\eqref{eq:hom_rel} holds.
The interpretation of eq.~\eqref{eq:hom_rel} is straightforward: each integration contour that encircles the singularity at $(X_{\infty}Y)=0$ as well as the poles of a subset of propagators can be replaced by a linear combination of integration contours that only encircle propagator poles. As we will see in what follows, and further in Section~\ref{sec:ibps}, this implies relations between different cuts of a given Feynman integral.

\subsection{Cut integrals associated to singularities of the second type}
\label{sec:cut_infty_rel}
The discussion in the previous sections allows us to extend the definition of cut integrals to singularities of the second type. In analogy with 
eq.~\eqref{eq:final_cut_def}, we define
\beq\label{eq:final_cut_def_2}
\cC_CI_n =  (2\pi i)^{-\lceil c/2\rceil}\,\int_{\tilde\Gamma_C}\varpi_n^D \mod i\pi\,, \qquad C\subseteq \{1,\ldots,n,\infty\}\,.
\eeq
In the special case where $\infty\notin C$, we can use the residue theorem, and we  recover the usual definition of cut integrals,
\beq\label{eq:cut_Res_compact}
\cC_{C}I_n=(2\pi i)^{-\lceil c/2\rceil}\,\int_{\tilde\Gamma_C}\varpi_n^D = (2\pi i)^{-\lceil c/2\rceil}\,\int_{\delta_C\tilde S_C}\varpi_n^D = (2\pi i)^{\lfloor c/2\rfloor}\,\int_{\tilde S_C}\textrm{Res}_C[\varpi_n^D]\,.
\eeq
If $\infty \in C$, we can use eq.~\eqref{eq:hom_rel} and write the integral $\cC_CI_n$ in terms of cut integrals that can be evaluated using the residue theorem. Note that many of the integrals resulting from the integration over the contour on the right-hand side of eq.~\eqref{eq:hom_rel} can be dropped, because upon applying the residue theorem we see that these terms are proportional to additional powers of $2\pi i$. 
In general, we can always express cut integrals associated to singularities of the second type in terms of ordinary cut integrals. We find
\begin{itemize}
\item for $|C|$ even,
\beq\label{eq:homo_even}
\cC_{\infty C}I_n = \sum_{i\in [n]\setminus C} \cC_{Ci}I_n + \sum_{\substack{i,j\in [n]\setminus C\\ i<j}}\cC_{C ij}I_n\mod i\pi\,.
\eeq
\item for $|C|$ odd,
\beq\label{eq:homo_odd}
\cC_{\infty C}I_n = -2\cC_{C}I_n-\sum_{i\in [n]\setminus C} \cC_{Ci}I_n \mod i\pi\,.
\eeq
\end{itemize}

Although cut integrals associated to singularities of the second type can be expressed in terms of ordinary cut integrals, it can be interesting to try to evaluate them directly. Unlike for singularities of the first type, the integration contour crosses a cut, and we cannot evaluate the integral in terms of residues. Instead, the integral reduces to an integral over the discontinuity across this cut~\cite{PhamBook}.
More precisely, one finds,
\beq\bsp\label{eq:cut_infty}
\cC_{\infty C}I_n &\,= (2\pi i)^{\lfloor (c+1)/2\rfloor}\, \int_{\partial_C \tilde E_{\infty C}}\textrm{Disc}_{{\infty}}\textrm{Res}_C[\varpi_n^D] \mod i\pi\\
&\,= -2\eps\,(2\pi i)^{\lfloor (c+1)/2\rfloor}\, \int_{\partial_C \tilde E_{\infty C}}\textrm{Res}_C[\varpi_n^D]\mod i\pi\,,
\esp\eeq
where the discontinuity operator is defined as the difference of the value of the function before and after analytic continuation, normalized by $(2\pi i)$, 
\beq\bsp
\textrm{Disc}_{\infty}(X_\infty Y)^{2\eps-x_n} &\,\equiv \frac{1}{2\pi i}\left( (X_\infty Y)^{2\eps-x_n}- [e^{2\pi i}(X_\infty Y)]^{2\eps-x_n}\right)\\
&\, =-2\eps\,(X_\infty Y)^{2\eps-x_n} \mod i\pi\,.
\esp\eeq
The integration contour in eq.~\eqref{eq:cut_infty} is defined as follows: $\tilde E_{\infty C}$ is the \emph{vanishing cell} associated to the pinch, which in the compactified picture can be identified with the cell cut out\footnote{In general, there is more than one such cell, but only one of them vanishes as we approach the pinch singularity.} by the hyperplanes $P_{j}$, $j\in C\cup\{\infty\}$. By definition the boundary of the vanishing cell $\tilde E_{\infty C}$ is contained in the union of the singular surfaces $\tilde S_{j}$, $j\in C\cup\{\infty\}$. The operator $\partial_C$ associates to $\tilde E_{\infty C}$ the part of the boundary contained in the union of the singular surfaces $\tilde S_j$, $j\in C$. Although a direct computation of the integral in eq.~\eqref{eq:cut_infty} may be hard to set up, we can equate eq.~\eqref{eq:cut_infty} with eqs.~\eqref{eq:homo_even} and~\eqref{eq:homo_odd} to obtain interesting relations among cut integrals. This will be analyzed in detail in Section~\ref{sec:ibps}.

% !TEX root = main-cuts.tex

\section{Cut and uncut integrals as a single class of parametric integrals}
\label{sec:loop-cut-duality}

In this section, we argue that the compactification  presented in \refE{eq:compactified} reveals that one-loop Feynman integrals and their cuts belong to the same class of functions. 
Moreover, this class of functions can be easily written as parametric integrals, thus providing an alternative way of evaluating cut Feynman integrals. As in the previous section, we work in Euclidean kinematics.

\subsection{Cut and uncut integrals in projective space}

Let us define a class of functions by
\beq\label{eq:Q_n_def}
Q_{n}^D(X_1,\ldots,X_n,X_{0}) = \frac{(-1)^{n}\,e^{\gamma_E\eps}}{\pi^{D/2}}\int_{\Sigma}\frac{d^{D+2}Y\,\delta((YY))}{\textrm{Vol}(GL(1))}\,\frac{[-2(X_{0}Y)]^{n-D}}{[-2(X_1Y)]\ldots [-2(X_nY)]}\,,
\eeq
where the integration runs over the real quadric $(YY)=0$ in $\mathbb{CP}^{D+1}$. The definition~\eqref{eq:Q_n_def} is very reminiscent of eq.~\eqref{eq:compactified}, the only difference being that in eq.~\eqref{eq:Q_n_def} the point $X_0$ is generic and not restricted to be the point $X_{\infty}$ as defined in \refE{eq:infinityPoint} (in particular, we do not require $X_0$ to be lightlike).
It is clear that every one-loop Feynman integral is a special case of eq.~\eqref{eq:Q_n_def},
\beq
I^D_n(X_1,\ldots,X_n) = Q^D_n(X_1,\ldots,X_n,X_{\infty})\,.
\eeq
More generally, whenever $X_0$ is lightlike, i.e.\ $(X_0X_0)=0$, we can find an $SO(D+1,1)$ transformation that maps $X_0$ to $X_{\infty}$, so that $Q^D_n(X_1,\ldots,X_n,X_{0})$ evaluates to a one-loop Feynman integral.
We will now argue that every one-loop cut integral also evaluates to an integral of the type~\eqref{eq:Q_n_def}, albeit in a case where $X_0$ is not necessarily lightlike. 

Since we have already established that every one-loop cut integral associated to a Landau singularity of the second type can be written as a linear combination of cut integrals associated to singularities of the first type, we restrict the discussion to singularities of the first type. We therefore assume without loss of generality that the set of cut propagators is $C=[c]$.
More concretely, we wish to compute the cut integrals $\cC_{C}I_n$  in eq.~(\ref{eq:cut_Res_compact}) using the same projective space formulation. We start by computing the residue of the integrand where the propagators in $C$ are on shell. In order to do this, it is convenient to change variables to include the set of cut propagators. We do this in two steps: we first change variables to $a_i$ and $Y_{\bot}$, defined by
\beq\label{eq:Ybot}
Y = Y_{\bot} + \sum_{i=1}^ca_iX_i\,,\quad \textrm{with }(X_iY_{\bot})=0 \textrm{ for }1\le i\le c\,,
\eeq
followed by a change of variables from $a_i$ to the cut propagators
\beq
D_i = (X_iY) = \sum_{j=1}^ca_j(X_iX_j)\,,\quad i\in C\,.
\eeq
The integration measure on $\mathbb{CP}^{D+1}$ takes the form
\beq
d^{D+2}Y = \frac{1}{\sqrt{Y_C}}\,d^cD_i\,d^{D-c+2}Y_{\bot}\,.
\eeq
After these changes of variables, the cut integral of eq.~(\ref{eq:cut_Res_compact}) is given by
\beq
\cC_CI_n^D = \frac{(-1)^{n}\,(2\pi i)^{\lfloor c/2\rfloor}\,e^{\gamma_E\eps}}{(-2)^c\,\pi^{D/2}\sqrt{Y_C}}\int_{\tilde{S}_C}\frac{d^{D-c+2}Y_{\bot}\,\delta((Y_\bot Y_\bot))}{\textrm{Vol}(GL(1))}
\frac{[-2(X_{\infty}Y_{\bot})]^{n-D}}
{[-2(X_{c+1}Y_{\bot})]\ldots [-2(X_{n}Y_{\bot})]},
\eeq
where the integration runs over the vanishing sphere $\tilde{S}_C = \Sigma\cap P_C$, defined by the equation $(Y_\bot Y_\bot)=0$ and we recall the result is valid modulo $i\pi$. Note that $\cC_CI_n^D$ is not a function of the scalar products of the points $X_i\in\mathbb{CP}^{D+1}$, $i>c$,
but rather of the scalar product of the projections of the $X_i$ onto the subspace $P_C$. To make this explicit we can write another projection, similar to \refE{eq:Ybot}, for $i> c$,
\begin{equation}\label{eq:X'}
	X_i=X'_{C,i}+\sum_{j=1}^c\alpha_{ij}X_j \,,\quad \textrm{with }(X_jX'_{C,i})=0 \textrm{ for }1\le j\le c\,,
\end{equation}
such that $X'_{C,i}\in P_C$. Then, using the fact that $(X_jY_\bot)=0$ for any $j\in C$ we obtain
\begin{align}
\nonumber
\cC_CI_n^D &\,= \frac{(-1)^{n}\,(2\pi i)^{\lfloor c/2\rfloor}\,e^{\gamma_E\eps}}{(-2)^c\,\pi^{D/2}\sqrt{Y_C}}\int_{\tilde{S}_C}\frac{d^{D-c+2}Y_{\bot}\,\delta((Y_\bot Y_\bot))}{\textrm{Vol}(GL(1))}
\,\frac{[-2(X'_{C,\infty}Y_{\bot})]^{(n-c)-(D-c)}}{[-2(X'_{C,c+1}Y_{\bot})]\ldots [-2(X'_{C,n}Y_{\bot})]}\\
\label{eq:cut-loop-duality}
&\,= \frac{2^{-c}\,(2\pi i)^{\lfloor c/2\rfloor}}{\pi^{c/2}\sqrt{Y_C}}\,Q_{n-c}^{D-c}(X'_{C,c+1},\ldots,X'_{C,n},X'_{C,\infty})\,.
\end{align}
We see that, up to an overall factor, one-loop cut integrals can be expressed in terms of the integrals defined in eq.~\eqref{eq:Q_n_def}. Unlike for uncut integrals, however, the `point at infinity' is in general not lightlike, $(X'_{C,\infty}X'_{C,\infty})\neq0$. The explicit form of the points $X'_{C,i}$ is given below.

The geometric picture underlying eq.~\eqref{eq:cut-loop-duality} is clear: if we want to compute the cut integral associated to a set $C$ of propagators, we have to intersect the quadric $\Sigma$ and the hyperplanes $P_i$, $i>c$, by the subspace $P_C$ where the propagators are on shell. In the subspace $P_C$, the new integration region is the vanishing sphere $\tilde{S}_C = \Sigma\cap P_C$, and the singularities are located on the hyperplanes $P_{Ci}\equiv P_i\cap P_C$ of $P_C$ defined by the equations $(X'_{C,i}Y_{\bot})=0$. In other words, and more loosely speaking, the operation of cutting a subset of propagators at one loop corresponds, geometrically, to the operation of intersection with the subspace $P_C$ where the propagators are on shell.

For \refE{eq:cut-loop-duality} to be useful in computing cut integrals, one needs to have an explicit form for the points $X'_{C,i}\in P_C$.
We can determine the coefficients $\alpha_{ij}$ in \refE{eq:X'} in terms of the scalar products $(X_iX_j)$, for $1\leq j\leq c$.
Seen as points in the ambient space $\mathbb{CP}^{D+1}$, it then follows that the $X'_{C,i}$ are given by
\beq\label{eq:x_prime_def}
X'_{C,i} = \frac{1}{(-1)^c Y_C}\,\det\left(\begin{array}{cccc}
(X_1X_1) & \ldots &(X_1X_c)& X_1\\
\vdots && \vdots&\vdots \\
(X_cX_1) & \ldots &(X_cX_c)& X_c\\
(X_iX_1) & \ldots &(X_iX_c)& X_i
\end{array}\right)\,.
\eeq
It is clear from this representation that $X'_{C,i}$ satisfies $(X_jX'_{C,i})=0$  for $1\le j\le c$, as $(X_jX'_{C,i})$ is the determinant of a matrix whose $j$-th and last column are equal.
From \refE{eq:X'}, we also find that $Y_{\bot}\in P_{Ci}$ if and only if $(X'_{C,i}Y_{\bot})=0$.
Finally, we note that scalar products involving the points $X'_{C,i}$ can be related to the scalar product between the points $X_i$. Indeed,
using $(X'_{C,i}X'_{C,j}) = (X_iX'_{C,j})$, one can easily check that
\beq\label{eq:proj_scalar_product}
(X'_{C,i}X'_{C,j}) = \frac{1}{(-1)^c Y_C}\,\det\left(\begin{array}{cccc}
(X_1X_1) & \ldots &(X_cX_1)& (X_iX_1)\\
\vdots && \vdots&\vdots \\
(X_1X_c) & \ldots &(X_cX_c)& (X_iX_c)\\
(X_1X_j) & \ldots &(X_cX_j)& (X_iX_j)
\end{array}\right)\,.
\eeq
With this expression and \refE{eq:compact_Gram}, we can rewrite all invariants on which 
the functions $Q_{n-c}^{D-c}(X'_{C,c+1},\ldots,X'_{C,n},X'_{C,\infty})$ depend in terms of Gram and modified Cayley determinants of the corresponding Feynman diagrams.

\subsection{Cut and uncut integrals as parametric integrals}

We now discuss how the integrals $Q_n^D$ introduced in the previous section can be written as parametric integrals. We follow closely the discussion of \cite{SimmonsDuffin:2012uy} where integrals of this form have been computed. As argued there, there is a unique conformal integral depending on a single point $X\in\mathbb{CP}^{D+1}$,
\begin{equation}\label{eq:confInt}
	I(X)=\int_{\Sigma}\frac{d^{D+2}Y\,\delta((Y Y))}{\textrm{Vol}(GL(1))}
	[-2(Y X)]^{-D}=\frac{\pi^{D/2}\Gamma(D/2)}{\Gamma(D)}
	[-(X X)]^{-D/2}\,.
\end{equation}

To see why this integral is relevant for the $Q_n^D$, we start from \refE{eq:Q_n_def} and introduce Feynman parameters to combine all propagators.\footnote{For a maximal cut, there is no remaining parametric integration to perform: using \refE{eq:confInt} in \refE{eq:cut-loop-duality} we directly reproduce the results in eqs.~\eqref{eq:MaxCutEven} and \eqref{eq:MaxCutOdd}.} It is then straightforward to get a parametric integral representation for the $Q_n^D$:
\begin{align}\label{eq:paramIntai}
	Q_n^D(X_1,\ldots,X_n,X_0)=
	\frac{(-1)^n\,e^{\gamma_E\eps}}{\pi^{D/2}}\frac{\Gamma(D)}{\Gamma(D-n)}
	\int [da]\, a_0^{D-n-1}\,I(\xi),
\end{align}
whith $I(\xi)$ as given in \refE{eq:confInt} and where we defined
\begin{equation}
	\xi\equiv\sum_{i=0}^na_iX_i\,,\qquad\quad
	\int [da]\equiv\left(\prod_{i=0}^n\int_0^\infty da_i\right)\,\delta(1-h(a))\,,
\end{equation}
with $h(a)=\sum_{i=0}^n h_i a_i$ such that the $h_i\geq 0$ are not all zero (see e.g.~\cite{Panzer:2015ida}). In particular, if we take 
$h_i=\sqrt{(X_iX_i)}$ and change variables to $b_i=h_i\,a_i$, then
\begin{align}\bsp
	&Q_n^D(X_1,\ldots,X_n,X_0)=\\
	&=\frac{(-1)^ne^{\gamma_E\eps}\Gamma(D/2)}{\Gamma(D-n)}\!
	\int [db]\left(\frac{b_0}{\sqrt{(X_0X_0)}}\right)^{D-n-1}\!
	\left(-\sum_{i=0}^n b_i^2-2\!\!\sum_{\substack{i, j=0 \\ i<j}}^n b_i b_j u_{ij}\right)^{-D/2}\!\!,
\esp\end{align}
where
\begin{equation}
	u_{ij}\equiv\frac{(X_iX_j)}{\sqrt{(X_iX_i)(X_jX_j)}}\,,\qquad\quad
	\int [db]\equiv\left(\prod_{i=0}^n\int_0^\infty \frac{db_i}{\sqrt{(X_iX_i)}}\right)
	\,\delta\left(1-\sum_{j=0}^nb_j\right)\,.
\end{equation}
The change of integration variables to the $b_i$ makes explicit that the result of the remaining parametric integrations is a function of the conformally invariant ratios $u_{ij}$. This change of variables is not defined if $(X_iX_i)=0$ for some $i$, as is the case, for instance, for an uncut integral for which $X_0=X_\infty$. However, the parametric representation of \refE{eq:paramIntai} is always well defined, and we will now see how it leads to the usual Feynman parameter representation of uncut Feynman integrals.

We take $h_0=0$ in eq.~\eqref{eq:paramIntai}, and rewrite it as
\begin{equation}
	Q_n^D(X_1,\ldots,X_n,X_0)=
	\frac{(-1)^n\,e^{\gamma_E\eps}\,\Gamma(D/2)}{\Gamma(D-n)}
	\int_0^\infty da_0\,a_0^{D-n-1}\int [da]_1\,\,[-(\xi \xi)]^{-D/2}\,,
\end{equation}
with
\begin{equation}
	\int [da]_1\equiv\left(\prod_{i=1}^n\int_0^\infty da_i\right)\,\delta(1-h_1(a))\,,\quad\qquad
	h_1(a)=\sum_{i=1}^n h_i a_i\,.
\end{equation}
We now set $X_0=X_\infty$, where $X_\infty$ is the lightlike point defined in \refE{eq:infinityPoint}, and recall that
\begin{equation}
	(X_\infty X_i)=-\frac{1}{2}\,,\quad\quad
	(X_iX_j)=-\frac{1}{2}\left((q_i^E-q_j^E)^2+m_i^2+m_j^2\right)
	\equiv -\Upsilon^E_{ij}\,,
\end{equation}
where $\Upsilon^E$ is the Cayley matrix, i.e., the matrix whose principal minors are the modified Cayley determinants $Y_{[i]}$, written in Euclidean kinematics. We thus find that
\begin{equation}\label{eq:temp14122017}
	(\xi \xi)=-a_0\sum_{i=1}^na_i-\sum_{i,j=1}^n a_i\,a_j\,\Upsilon^E_{ij}
	=-a_0\,\mathcal{U}_n-\mathcal{F}_n\,,
\end{equation}
where $\mathcal{U}_n$ and $\mathcal{F}_n$ are respectively the first and second Symanzik polynomials associated with $n$-point one-loop graphs \cite{Bogner:2010kv} we also introduced in Appendix \ref{app:Landau}. Because \refE{eq:temp14122017} is linear and not quadratic in $a_0$, the integral over $a_0$ can be trivially performed to obtain
\begin{equation}
	Q_n^D(X_1,\ldots,X_n,X_\infty)=(-1)^n\,e^{\gamma_E\eps}\,\Gamma(n-D/2)
	\int [da]_1\frac{\mathcal{U}^{n-D}_n}{\mathcal{F}_n^{n-D/2}}\,,
\end{equation}
which is the usual Feynman parameter representation of the one-loop Feynman integral $I_n^D$. For completely generic kinematics, we can again change variables from the $a_i$ to the $b_i$ in the remaining integrations.

Thus we have seen that the functions $Q_n^D$ defined in eq.~\eqref{eq:Q_n_def} provide a unified framework for studying one-loop integrals and their cuts. The $Q_n^D$ corresponding to either cut or uncut integrals can be easily written as parametric integrals. This observation is particularly useful when computing cut integrals, as it avoids the need to parametrize the loop momentum as was done in \refS{sec:parametrization}.
Finally, given this framework we should expect that there is also a unified way to describe the discontinuities of one-loop integrals, cut or uncut. In particular, it is well known that cuts of Feynman integrals are related to their discontinuities, and it would be interesting to understand how this observation generalizes when starting with cut Feynman integrals, and considering their cuts and discontinuities. This will be investigated in detail in the next section.

% !TEX root = main-cuts.tex

\section{Cut integrals and discontinuities}
\label{sec:cuts_&_discontinuities}

Cut integrals are closely connected to discontinuities. In this section, we investigate this statement, taking Picard-Lefschetz theory as our mathematical foundation. We discuss its application to cut Feynman integrals and discuss the interpretation of discontinuities, iterated discontinuities, the crossing of branch cuts, and leading singularities.

\subsection{Discontinuities}
\label{sec:discontinuities}

It has been known since the early days of quantum field theory that Feynman integrals are not single-valued functions of the external kinematics. Rather, they have discontinuities that are associated to internal propagators going on shell. The values of these discontinuities are computed by cut integrals~\cite{Cutkosky:1960sp}. This idea can be described rigorously through the Picard-Lefschetz theorem and the multivariate residue theorem of Section~\ref{sec:residues}~\cite{PhamCompact,Teplitz,PhamBook}. Although these results have long been known in the mathematical physics literature, we feel that many of these ideas have slipped into oblivion.  Therefore, in this section we  present  a short overview of Picard-Lefschetz theory and how it relates to the cuts and discontinuities of Feynman integrals. We do not aim at mathematical rigor here (see, e.g., the recent ref.~\cite{Bloch:2015efx} for a rigorous proof of the material reviewed in this section), but we rather present the general ideas.

Consider a function $f(t)$ defined by an integral
\beq\label{eq:int_example}
f(t) = \int_{\Gamma}\omega_t\,.
\eeq
Assume that the integrand has singularities on  hypersurfaces $S_j$, $1\le j\le n_s$, and that the boundary $\partial\Gamma$ is contained in the hypersurfaces $B_k$, $1\le k\le n_b$. We assume that the singular surfaces $S_j$ and the boundaries $B_k$ are given by the equations
\beq
s_j(z,t) = 0 {\rm~~and~~} b_k(z,t) = 0\,.
\eeq
The singularities and branch points of $f$ arise from pinch and/or endpoint singularities of the integral~\eqref{eq:int_example}. 
A necessary condition for a pinch or endpoint singularity to occur is that there exist a set of $\alpha_j$ and $\beta_k$, not all zero, such that~\cite{Landau:1959fi}
\begin{align}\label{eq:Landau_1}
&\alpha_j\,s_j(z,t) = 0\,, \quad \beta_k\,b_k(z,t) = 0\,, \quad 1\le j\le n_s\,,\quad 1\le k\le n_b\,,\\
\label{eq:Landau_2}
&\frac{\partial}{\partial z_i}\left[\sum_{j=1}^{n_s}\alpha_j\,s_j(z,t) + \sum_{k=1}^{n_b}\beta_k\,b_k(z,t)\right] = 0\,.
\end{align}
Equations~\eqref{eq:Landau_1} and~\eqref{eq:Landau_2} are the first and second Landau conditions, which were  already presented in \refS{sec:polytopes} in a different representation.
The representation in equations~\eqref{eq:Landau_1} and~\eqref{eq:Landau_2} makes clear that the Landau conditions treat the singular surfaces $S_j$ and the boundaries $B_k$ on the same footing. We therefore only discuss the case of singularities, from which the extension to boundaries is straightforward.

The set of values of $t$ for which a nontrivial solution to the Landau conditions exists is called the \emph{Landau variety} $L$ of $f$. In practice it is convenient to decompose $L$ into a union of components $L_C$, where $C$ is the subset of $\alpha_j$ that do not vanish for this particular solution of the Landau conditions.

Assume now that we start from a point $t\notin L$ and we analytically continue $t$ along a small loop around $L_C$. Since the integration contour $\Gamma$ may vary with $t$ (e.g., we may need to deform the integration contour in order to avoid a singularity that approaches it), the integration contour may have changed into a new contour $\Gamma^{(C)}$ at the end of the analytic continuation. The change in the integration contour is computed by the Picard-Lefschetz theorem,
\beq
\Gamma^{(C)} - \Gamma = N_C\,\delta_C\sigma\,,
\eeq
where $N_C$ is an integer whose value is not important for this introductory discussion, $\delta_C$ is the iterated Leray coboundary associated to the singular surfaces $S_j$ for $j\in C$, and $\sigma$ is a cycle in $S_C= \bigcap_{j\in C}S_j$ with the topology of a sphere, which coincides with the vanishing sphere. The discontinuity of the function $f$ around the Landau variety $L_C$ is then defined as the difference between before and after analytic continuation,
\beq\label{eq:Var_def}
\textrm{Disc}_{L_C}\,f \equiv \int_{\Gamma}\omega_t - \int_{\Gamma^{(C)}}\omega_t = -N_C\,\int_{\delta\sigma}\omega_t = -(2\pi i)^c \,N_C\,\int_{\sigma}\textrm{Res}_C[\omega_t]\,,
\eeq
where the last step follows from the residue theorem~\eqref{eq:residue_thm} in the case where the integrand has poles but no branch cuts.

As an example, consider the function $f$ defined by
\beq
f(a,b) = \int_{0}^\infty\frac{dz}{(z-a)(z-b)}\,,
\eeq
where we assume for simplicity that $a\neq b$ are non-positive complex numbers. The Landau conditions for $f$ are then
\beq\bsp
&\alpha_1\,(z-a)=0\,,\quad\alpha_2\,(z-b)=0\,,\quad\beta_1\,z=0\,,\quad \alpha_1+\alpha_2+\beta_1=0\,.
\esp\eeq
It is easy to check that for $a\neq b$, $\alpha_1$ and $\alpha_2$ cannot be nonzero simultaneously. We assume from now on that $\alpha_2=0$. In that case, the only possibility for a nontrivial solution is to have both $\alpha_1$ and $\beta_1$ nonzero, which implies $a=0$. The analysis for $\alpha_1=0$ is similar, giving $b=0$. Hence, the Landau variety is the union of two connected components $L=L_a\cup L_b$, where each $L_x$ is the isolated point defined by $x=0$. 

Let us now analytically continue $a$ along a small loop around $a=0$, while keeping $b$ fixed. The change in the function $f$ is given by eq.~\eqref{eq:Var_def}. Since the singular surfaces consist of isolated points, the vanishing sphere $\sigma$ is the isolated point $z=a$, and the corresponding vanishing cycle $\delta \sigma$ is a small circle around $a$. We then find
\beq
\textrm{Disc}_{L_a}\,f = -N_a\,\int_{\delta\sigma}\frac{dz}{(z-a)(z-b)} = -2\pi i \,N_a\,\int_{z=a}\textrm{Res}_{z=a}[\omega] = -\frac{2\pi i\,N_a}{a-b}\,.
\eeq
It is easy to check that this is indeed the correct result (with $N_a=1$ in this case).

%%%%%%%%%%%%%%%%%%%%%%%%%
\subsection{Discontinuities of one-loop integrals}
\label{sec:disc_1-loop}
In this section we review how the ideas from the previous section apply to Feynman integrals and how, as a consequence, cut integrals can be interpreted as discontinuities of Feynman integrals. Specializing to one-loop integrals, the Landau conditions~\eqref{eq:Landau_1} and~\eqref{eq:Landau_2} reduce to the usual Landau conditions~\eqref{eq:landau1} and~\eqref{eq:landau2} for one-loop integrals (see also Appendix~\ref{app:Landau}). We have already seen that the solutions to the Landau conditions for one-loop integrals can be classified according to whether the modified Cayley or Gram determinant vanishes.
Hence, the Landau variety of a one-loop Feynman integral takes the form\footnote{Eq.~\eqref{eq:fullLandau} does  not include $L_\infty$, because it would correspond to the vanishing of $\det(X_{\infty}X_{\infty})$, but since this determinant vanishes identically (i.e., independently of the external kinematics), it does not define a surface in the space of the external kinematics.}
\beq\label{eq:fullLandau}
L=\bigcup_{\substack{\emptyset\subset C\subseteq [n]}}L_C\cup L_{\infty C}\,,
\eeq
where 
\beq\bsp
L_C &\,=
\{(q_j,m_j^2)_{1\le j\le n} : Y_C = 0\}\,,\\
L_{C\infty} &\,= \{(q_j,m_j^2)_{1\le j\le n}  : \Gram_C = 0\}\,.
\esp\eeq

In the rest of this section, we concentrate on the Landau singularities of the first type; Landau singularities of the second type are then related as discussed in Section~\ref{sec:cut_infty_rel}. We want to compute the discontinuity of $I_n$ when the external kinematics is analytically continued around $L_C$ for some $C\subseteq [n]$ (while not encircling any other Landau variety). From eq.~\eqref{eq:Var_def}, we obtain
\beq\bsp\label{eq:disc_cut}
\textrm{Disc}_CI_n&\, \equiv \,(2\pi i)^{-\lceil c/2\rceil}\,\textrm{Disc}_{L_C}I_n \\
&\,= -\,N_C\,(2\pi i)^{\lfloor c/2\rfloor}\, \int_{\tilde{S}_C}\textrm{Res}_C[\varpi_n^D]\\
&\,=-N_C\,\cC_CI_n \mod\, i\pi\,,
\esp\eeq
where we recall that we have defined the integrals $\cC_CI_n$ only modulo $i\pi$. A similar relation holds for higher loop integrals~\cite{Cutkosky:1960sp,PhamCompact,Teplitz,Bloch:2015efx}. Note that not every discontinuity is necessarily present on every Riemann sheet of $I_n$. The integer $N_C$ can easily be determined in the compactified picture for one-loop integrals. In fact, $N_C$ is related to the \emph{intersection index} of the vanishing cell $\tilde E_C$ and the integration contour $\Sigma$,
\beq
N_C = [\Sigma,\tilde E_C]\,.
\eeq
In ref.~\cite{PhamInTeplitz} it was shown that $[\Sigma,\tilde E_C]=\pm1$.

%%%%%%%%%%%%%%%%%%%%%%%%%%%%%%%%%%%%%%%%%%
\subsection{Iterated discontinuities}

As reviewed in the beginning of this section, cuts compute discontinuities, i.e.\ the change of a Feynman integral as the external kinematics and/or  masses are analytically continued around a Landau variety $L_C$. In this subsection, we address the  case of iterated discontinuities, i.e.\ performing successive analytic continuations around different Landau varieties. 
In Section~\ref{sec:loop-cut-duality} we argued that at one loop, both cut and uncut integrals can be expressed via the same class of functions $Q_n^D$, and that the operation of cutting, and thus of analytic continuation, has a very simple geometric interpretation in terms of intersections of hyperplanes. Since the operation of taking intersections is associative and commutative, it is natural to expect that iterated discontinuities follow a similar simple pattern. We now show that this is indeed the case.

Assume that we have computed the discontinuity of a one-loop integral around the Landau variety $L_C$ (cf. eq.~\eqref{eq:disc_cut}). The resulting discontinuity function is still multi-valued, and our goal is to compute the discontinuities of the discontinuity function\footnote{We had defined our cut integrals modulo $i\pi$. Throughout this section we sightly lift this restriction and study how the cuts change under analytic continuation.} $\textrm{Disc}_C I_n^D$. We only discuss singularities of the first type, because all discontinuities around Landau varieties associated to singularities of the second type can be expressed in terms of those of the first type. Our first goal is to describe the Landau varieties of $\textrm{Disc}_C I_n^D$. It is natural to expect that the Landau varieties of $\textrm{Disc}_C I_n^D$ are contained in those of $I_n^D$, and indeed this result follows from the unified framework for cut and uncut integrals dicussed in Section~\ref{sec:loop-cut-duality}.
Starting from eq.~\eqref{eq:cut-loop-duality}, it is easy to check that the Landau varieties associated to the singularities of the first type, seen as a function of the points $X'_{C,i}$, $i>c$, are classified by subsets $C'\subseteq \{c+1,\ldots,n\}$. More precisely, they are given by
\beq
L_{C,C'} : Y_{C,C'} \equiv \det(X'_{C,i}X'_{C,j})_{i,j\in C'}=0\,.
\eeq
Since the points $X'_{C,i}$ are functions of the external kinematics (they are functions of the points $X_i$), we can express $Y_{C,C'}$ in terms of the points $X_i$. We show at the end of this section that the following identity holds:
\beq\label{eq:iterated_Landau}
Y_{C,C'} = \frac{Y_{C\cup C'}}{Y_C}\,,\quad C,C'\subseteq [n]\,,\quad C\cap C'=\emptyset\,.
\eeq
Before we present the proof of this relation, let us explore some of its consequences.
The previous equation implies that $L_{C,C'}\subset L_{C\cup C'}$, and so an iterated discontinuity around $L_C$ and $L_{C,C'}$ must be related to analytic continuation around $L_{C\cup C'}$. If we denote by $\textrm{Disc}_{C,C'}$ the operation of taking the discontinuity of $\textrm{Disc}_{C}I_n^D$ around the Landau variety $L_{C,C'}$, we find,
\begin{align}\label{eq:iterated_disc}
\textrm{Disc}_{C,C'}\textrm{Disc}_{C}I_n^D &\,= -N_C\,\textrm{Disc}_{C,C'}\cC_CI_n\\
\nonumber&= -N_C\,\frac{2^{-c}\,(2\pi i)^{\lfloor c/2\rfloor}}{\pi^{c/2}\sqrt{\mu^c\,Y_C}}\,\textrm{Disc}_{C,C'}Q_{n-c}^{D-c}(X'_{C,c+1},\ldots,X'_{C,n},X'_{C,\infty})\\
\nonumber&= N_C\,N_{C'}\frac{2^{-c-c'}(2\pi i)^{\lfloor c/2\rfloor+\lfloor c'/2\rfloor}}{\pi^{(c+c')/2}\sqrt{\mu^{c+c'}\,Y_C\,Y_{C,C'}}}Q_{n-c-c'}^{D-c-c'}(X''_{C,c+c'+1},\ldots,X''_{C,n},X''_{C,\infty})\\
\nonumber&= -\frac{N_C\,N_{C'}}{N_{C\cup C'}}\,(2\pi i)^{\lfloor c/2\rfloor+\lfloor c'/2\rfloor-\lfloor (c+c')/2\rfloor}\,\textrm{Disc}_{C\cup C'}I_n^D\,.
\end{align}
In other words, we see that, up to an overall constant numerical factor, iterated discontinuities around a Landau variety where an additional subset $C'$ of propagators is cut is equivalent to computing the discontinuity where all the propagators in $C\cup C'$ are on shell. This result is in agreement with the findings of ref.~\cite{Boyling}. Since the operation of taking unions is associative and commutative, we conclude that at one loop the operation of taking discontinuities must have the same properties. It is not clear, however, if this simple picture survives at higher loop orders~\cite{Boyling,Hierarchy}.

We now give the proof of eq.~\eqref{eq:iterated_Landau}. Consider two disjoint subsets $C$ and $C'$ of $[n]$. Before we discuss the details of the proof, we mention the following simple geometric observation. The intersection with the subspace $P_{C\cup C'} = P_C\cap P_{C'}$ determines a set of points $X'_{C\cup C',i} \in P_{C\cup C'}$, $i\in[n]\setminus(C\cup C')$. We can compute the same intersection sequentially: We first compute the intersection with the subspace $P_C$, which determines a set of points $X'_{C,i}\in P_C$, $i\in [n]\setminus C$, and we next compute the intersection with $P_{C'}$, which determines the points $X''_{C',C,i}\in P_{C\cup C'}$, $i\in[n]\setminus(C\cup C')$. Obviously, the two sets of points must agree, $X''_{C',C,i} = X'_{C\cup C',i}\in P_{C\cup C'}$, 
and eq.~\eqref{eq:proj_scalar_product} then implies
\beq
\frac{Y_{C\cup C'i}}{Y_{C\cup C'}} = (X'_{C\cup C',i}X'_{C\cup C',i}) = (X''_{C',C,i}X''_{C',C,i}) = \frac{Y_{C, C'i}}{Y_{C, C'}} \,,
\eeq
where we have used the shorthand $C'i\equiv C'\cup\{i\}$. The previous relation can be cast in the form
\beq
Y_{C,C'i} = \frac{Y_{C\cup C'i}}{Y_{C\cup C'}}\,Y_{C, C'}\,.
\eeq
The proof of eq.~\eqref{eq:iterated_Landau} now proceeds recursively in the number of elements of $C'$. 
If $C'$ contains a single element, say $C'=\{i\}$, eq.~\eqref{eq:iterated_Landau} reduces to eq.~\eqref{eq:proj_scalar_product} with $i=j$, and so eq.~\eqref{eq:iterated_Landau} holds if $C'$ consists of a single element. If $C'$ contains more than one element, say $C' = C''i$, $C''\neq\emptyset$, then eq.~\eqref{eq:iterated_Landau} applies inductively to $Y_{C, C''}$, and we find
\beq
Y_{C,C'} =Y_{C,C''i} = \frac{Y_{C\cup C''i}}{Y_{C\cup C''}}\,Y_{C, C''} = \frac{Y_{C\cup C''i}}{Y_{C\cup C''}}\,\frac{Y_{C\cup C''}}{Y_{C}} = \frac{Y_{C\cup C'}}{Y_{C}}\,,
\eeq
and so eq.~\eqref{eq:iterated_Landau} is proven.

This analysis requires that none of the modified Cayley determinants appearing in the formulas above vanish. As such, it is valid for generic kinematics and does not apply to cases such as the vanishing cuts listed in \refS{sec:vanishing}. For example, the nonvanishing quadruple cut of the zero-mass box cannot be derived as an iterated discontinuity if the first cut gives zero.

%%%%%%%%%%%%%%%%%%%%%%%%%%%%%%%%%%%%%%%%%%

\subsection{Unitarity cuts and discontinuities in physical channels}
\label{sec:cut_disc}

The previous definition of discontinuity is related to a notion encountered frequently in the physics literature, namely the discontinuity across a branch cut, defined as the difference of the values of a function $f$ as one approaches the real axis from opposite sides,
\beq\label{eq:disc_def}
\textrm{disc}_{x}\,f(x+i0) \equiv \frac{1}{2\pi i}\lim_{\eta\to0}\Big[f(x+ i\eta)-f(x- i\eta)\Big]\,,\quad x\in \mathbb{R}\,.
\eeq
In this section we explain how this notion of discontinuity is related to the analytic continuation discussed in Section~\ref{sec:disc_1-loop}.

Consider a function $f$ that is analytic on some open domain $D$ in the upper half-plane, whose closure contains an interval $I$ of the real axis on which $f$ is real and continuous. In the following we will always assume that $D$ is the upper half-plane. The Schwarz reflection principle then implies that $f$ can be analytically continued in a unique way to the domain $\overline{D}$ in the lower half-plane by $f(\bar{z})=\overline{f(z)}$. This is equivalent to
\beq
\textrm{Re}f(\bar{z}) = \textrm{Re}f(z) {\rm~~and~~}\textrm{Im}f(\bar{z}) = -\textrm{Im}f(z)\,.
\eeq
In other words, while the real part of $f$ is well-defined on the real axis, the imaginary part is discontinuous
and eq.~\eqref{eq:disc_def} reduces to
\beq
\textrm{disc}_{x}\,f(x+i0) = i\,\textrm{disc}_{x}\,\textrm{Im}f(x+i0) = 2i\,\textrm{Im}f(x+i0) = -2i\,\textrm{Im}f(x-i0)\,,\quad x\in \mathbb{R}\,.
\eeq
Let us now consider a function that has a branch point at the origin on the real axis. We assume that $f(x)$ is real for $x<0$, and that $f$ has a branch cut on the positive real axis. For every point $z$ in the upper half-plane there is a path $\gamma$ that connects it to $\bar{z}$ without crossing the branch cut. We can thus interpret $f(x-i0)$ as the result of analytically continuing $f(x+i0)$ along the path $\gamma$:
\begin{itemize}
\item If $x<0$, then this path can be chosen as the straight line from $x+i0$ to $x-i0$, and we obtain
\beq
\textrm{disc}_{x}f(x) = i\,\textrm{disc}_{x}\textrm{Im}f(x)  = 0\,.
\eeq
\item If $x > 0$, then this path winds once in the positive direction around the branch point at $x=0$. In other words, $f(x-i0)$ is obtained from $f(x+i0)$ by analytic continuation around the branch point, and so we find
\beq
\textrm{disc}_{x}f(x) = i\,\textrm{disc}_{x}\textrm{Im}f(x)  = \frac{1}{2\pi}\textrm{Disc}_{x=0}\textrm{Im}f(x) = \frac{1}{2\pi}\textrm{Im}\,\textrm{Disc}_{x=0}f(x)\,,
\eeq
where the last step follows form the fact that analytic continuation commutes with the operation of taking the imaginary part.
\end{itemize}
Combining the two cases, we see that
\beq\label{eq:disc_Im_Disc}
\textrm{disc}_{x}f(x+i0) = \frac{1}{2\pi}\,\theta(-x)\,\textrm{Im}\,\textrm{Disc}_{x=0}f(x)\,,
\eeq
where $\theta(x)$ denotes the Heaviside step function.
We see that the two notions of discontinuities carry equivalent information, at least in the case where the branch point lies on the real axis.

Let us now discuss how these concepts apply to one-loop Feynman integrals. It is well known that one-loop Feynman integrals are real in the Euclidean region where all masses are positive and all consecutive Mandelstam invariants are negative. Hence, one-loop Feynman integrals satisfy the Schwarz reflection principle. Feynman integrals have branch points on the real axis which correspond to thresholds and masses becoming negative.
Let us discuss the discontinuity in the channel $Q^2=(q_i-q_j)^2$.
It follows from the `cutting rules' of ref.~\cite{tHooft:1973pz,Veltman:1994wz} that the discontinuities in a given channel can be computed by replacing a subset of propagators by $\delta$ functions and fixing the energy flow through these cut propagators. This subset is simply the set of propagators that disconnects the graph into two disjoint graphs, where the total external momentum flowing through the cut lines into each subgraph is equal to $Q^2$. At one loop, every subset of two edges disconnects the graph, and so we obtain the well-known result that the discontinuity in a channel $Q^2$ is computed by the integral where two propagators are `cut', i.e., replaced by  $\delta$ functions. Similarly, it is known that one-propagator cuts compute the discontinuity in the mass of the cut propagator~\cite{Abreu:2015zaa}.

Let us see how these well-known results can be recovered from the more general notion of discontinuity reviewed at the beginning of the section.
We discuss the discontinuity in a channel, $Q^2 = (q_i-q_j)^2$. Consider the Landau variety defined by
\beq
Y_{ij} = -\frac{1}{4}\,\left[Q^2 - (m_i+m_j)^2\right]\,\left[Q^2 - (m_i-m_j)^2\right] = 0\,.
\eeq
The zeroes of $Y_{ij}$ are the threshold and pseudothreshold~\cite{SMatrix}. Assume that all other scales are held fixed. In the complex $Q^2$-plane, the Landau variety $L_{ij}=\{Q^2\in\mathbb{C}\,|\, Y_{ij}=0\}$ consists of two isolated points. We know that the amplitude has a branch cut along the positive real line starting from the physical threshold $Q_0^2 \equiv (m_i+m_j)^2$. Using eq.~\eqref{eq:disc_cut} and~\eqref{eq:disc_Im_Disc}, we find that
\beq
\textrm{disc}_{Q^2}I_n(Q^2) = \frac{1}{2\pi}\,\theta\!\left(Q^2-Q_0^2\right)\,\textrm{Im}\,\textrm{Disc}_{L_{ij}}I_n = -{N_{ij}}\,\theta\!\left(Q^2-Q_0^2\right)\,\textrm{Re}\,\cC_{ij}I_n\,,
\eeq
in agreement with the physics literature.

%%%%%%%%%%%%%%%%%%%%%%%%%%%%%%%%%%%%%%%%

\subsection{Leading singularities}
\label{sec:leading_singularities}

In Section \ref{sec:max_cut} we presented explicit results for maximal cuts of one loop integrals. A closely related notion is that of {\em leading singularities}. If we restrict the discussion to one-loop integrals, then the leading singularity is defined as the residue of the integrand at a global pole~\cite{Cachazo:2008vp}. This notion, however, only makes sense in integer dimensions. The residue is particularly simple to evaluate
in the compactified picture in $D=D_n$ dimensions ($D_n$ is defined in \refE{eq:defDn}). We consider separately the cases where $n$ is even or odd. We recall from the discussion of \refS{sec:mvresidues} that residues are a property of the integrand, and not of the integral. We thus compute the leading singularity as the global pole of the differential form $\varpi_n^{D_n}$ defined in \refE{eq:compactified} taken at $\eps=0$.
\begin{itemize}
\item If $n$ is even, then there is a global pole at $(X_iY)=0$, $1\le i\le n$. The residue at this pole evaluates to
\beq\label{eq:cL_J_n_even}
\LS[\varpi_n^{D_n}]=\pm \,\frac{(4\pi)^{-n/2}}{\sqrt{Y_{[n]}}}\,,\quad n \textrm{ even}.
\eeq
\item If $n$ is odd, then there is a global pole at $(X_{\infty}Y)=(X_iY)=0$, $1\le i\le n$. The computation is identical to the previous case, and using eq.~\eqref{eq:compact_Gram}, the residue at this pole evaluates to
\beq\label{eq:cL_J_n_odd}
\LS[\varpi_n^{D_n}]=\pm \,2i\frac{(4\pi)^{-(n+1)/2}}{\sqrt{\Gram_{[n]}}}\,,\quad n \textrm{ odd}.
\eeq
\end{itemize}

%\Einan{Added eq. refs. to below (\ref{eq:MaxCutEven}) and (\ref{eq:MaxCutOdd})}
Since one-loop leading singularities are residues at global poles, it is natural to expect that leading singularities are related to cut integrals where we integrate over the corresponding residues. 
We define\footnote{For cases where the maximal cut vanishes, which are strictly speaking not part of our basis as they are reducible to lower point functions, $j_n$ is defined by the expression in terms of determinants, and not as the $\eps\to0$ limit of the cut. }
\beq\label{eq:LS_J_n}
j_n\equiv \lim_{\eps\to0}\cC_{[n]}\tildeJ_n = \left\{\begin{array}{ll}
\displaystyle 2^{1-n/2}i^{n/2}\,/\sqrt{Y_{[n]}}\,, &\qquad \textrm{ for } n \textrm{ even}\,,\\ \,\\
\displaystyle 2^{(1-n)/2}i^{(n-1)/2}\,/\sqrt{\Gram_{[n]}}\,, & \qquad \textrm{ for } n \textrm{ odd}\,.
\end{array}\right.
\eeq
where we used the explicit results in (\ref{eq:MaxCutEven}) and (\ref{eq:MaxCutOdd}).
We then find the following relation between leading singularities and one-loop cut integrals:
\begin{itemize}
\item If $n$ is even,
\beq
j_n = \pm2(2\pi i)^{n/2}\,\LS[\varpi_n^{D_n}] \,.
\eeq
\item If $n$ is odd, using eq.~\eqref{eq:homo_odd},
\beq
\lim_{\eps\to 0}\cC_{\infty[n]}\tildeJ_n = -2\,j_n = \pm2\,(2\pi i)^{(n+1)/2}\,\LS[\varpi_n^{D_n}] \,.
\eeq
\end{itemize}
We see that the leading singularity and the maximal cut agree only up to a factor of 2 (the powers of $ 2\pi i$ on the right-hand side results from our normalization of cut integrals, see eq.~\eqref{eq:cutdef-det}). This factor is significant. Its origin  can be traced back to the vanishing sphere. 
Indeed, for $n$ even, the vanishing sphere is the intersection of $n$ spheres of dimension $n-1$, and for $n$ odd, the intersection of $(n+1)$ spheres of dimension $n$: in both cases the vanishing sphere reduces to two isolated points, which are precisely the two global poles. The integral over the vanishing sphere therefore reduces to the difference of the values of the residues at these global points. Since the residues differ by a sign, their difference leads to a factor of two. Equivalently, for the maximal cut the angular integral in eq.~\eqref{eq:cutdef-det} is trivial and contributes only the volume of the $0$-sphere, which is 2:
\beq
\lim_{\eps\to 0}\int d\Omega_{-2\eps} = \lim_{\eps\to0}\frac{2\pi^{(1-\eps)/2}}{\Gamma\left(\frac{1-\eps}{2}\right)} = 2\,.
\eeq
As a consequence, we emphasize that leading singularities should \emph{not} generally be interpreted as discontinuities! The former are a property of the integrand, the latter of the integral. While the difference amounts to a simple factor of 2 at one loop, this difference may be more pronounced beyond one loop where the structure of the global poles is more complicated~\cite{Kosower:2011ty,CaronHuot:2012ab,Sogaard:2014jla} and where there is no easy way to translate between the individual residues which define the leading singularities and the integrals over the vanishing spheres that define the discontinuities.

% !TEX root = main-cuts.tex

\section{Linear relations among cut integrals}
\label{sec:ibps}
The purpose of this section is to analyze linear relations among one-loop cut integrals.
It is well known that (uncut) Feynman integrals satisfy linear relations among themselves. A particular subset of such relations are the dimensional-shift identities~\cite{Bern:1992em,Tarasov:1996br,Lee:2009dh}, which relate integrals in $D$ and $D\pm2$ dimensions, and integration-by-parts (IBP) identities~\cite{Tkachov:1981wb,Chetyrkin:1981qh}, which are recursion relations in the powers of the propagators in a fixed space-time dimension. The origin of IBP identities lies in the fact that integrals of total derivatives vanish in dimensional regularization,
\beq\label{eq:int_tot_der}
\int d^Dk\,\frac{\partial}{\partial k^\mu}\Big(\ldots\Big)=0\,,
\eeq
which is a consequence of the invariance of integrals in dimensional regularization under infinitesimal linear changes of variables~\cite{Lee:2008tj}.
Acting with the derivative on the propagators (and expressing the scalar products in the numerator in terms of denominators) results in a linear combination of integrals with the same propagator structure, but with different values for the exponents of the propagators (numerator factors can be interpreted as propagators raised to non positive powers). Currently, no linear identity between Feynman integrals is known that cannot be traced back to IBP or dimensional-shift identities, and conjecturally all linear relations among Feynman integrals arise in this way. Using IBP identities (and tensor reduction in the presence of tensor numerators), every one-loop integral can be reduced to a linear combination of integrals with unit powers of the propagators. Using dimensional-shift identities, we can choose integrals with a different number of propagators to lie in a different number of dimensions. As a consequence, one can show that the integrals $\tildeJ_n$ defined in eq.~\eqref{eq:tildeJ_n} form a basis for all one-loop integrals in $D=d-2\eps$ dimensions, where $d$ is a positive even number. 

To summarize, conjecturally there are precisely two types of relations among one-loop Feynman integrals, IBP identities and dimensional-shift identities, and we can use these relations to construct a basis for all one-loop integrals.
Since linear relations among one-loop integrals are so well understood, it is natural to ask what  relations exist among cut integrals. In the following we argue that there are two types of relations among one-loop cut integrals:
\begin{enumerate}
\item Linear relations among  integrals with different propagators but the same set of cut propagators: these relations can be traced back to IBP and dimensional shift identities and are entirely determined by the corresponding relations between the uncut integrals.
\item Linear relations among integrals with the same propagators but different sets of cut propagators: these relations can be traced back to relations between the generators of the homology group associated to one-loop integrals studied in Section~\ref{sec:homology}.
\end{enumerate}
These different types of relations will be described more extensively in the remainder of this section.
Although we cannot exclude at this point that other types of (linear) relations among cut integrals exist, we have performed an extensive search for linear relations of the aforementioned type on explicit results for cut integrals with up to four propagators. We have not found any relation beyond the ones described in this section, so we conjecture that these are the only relations among one-loop cut integrals.

\subsection{Relations among integrals with the same set of cut propagators}
Since cut integrals compute discontinuities (see Section~\ref{sec:discontinuities}), every linear relation among Feynman integrals can immediately be lifted to a relation among cut integrals. Indeed, assume that we are given linear relations among a set of Feynman integrals $I_k$, $\sum_k a_k\,I_k=0$. Then we can compute the discontinuity associated to the Landau variety $L_C$, and we obtain a relation among cut integrals. For example, in the case of a singularity of the first type, we have 
\beq
0 = \sum_ka_k\,\textrm{Disc}_C I_k =-\,N_C \sum_ka_k\,\cC_C I_k\mod i\pi\,.
\eeq
Hence, we immediately conclude that every IBP and dimensional-shift identity can be lifted to a linear relation where precisely the propagators in the set $C$ are cut. The only difference with the uncut case comes from the fact that integrals vanish if a cut propagator is absent or raised to a negative power. Note that this result is not specific to one loop, but it can easily be generalized to arbitrary loop order, because the connection between discontinuities and cut integrals extends beyond one loop~\cite{Cutkosky:1960sp,Froissart,PhamBook,PhamCompact,PhamInTeplitz,Bloch:2015efx}.

The previous result is closely connected to the {reverse-unitarity} approach to the computation of inclusive cross sections. In ref.~\cite{Anastasiou:2002yz,Anastasiou:2002wq,Anastasiou:2002qz,Anastasiou:2003yy,Anastasiou:2003ds}, it was argued that the same IBP identities hold for unitarity cuts of the loop integral, i.e., cuts where the underlying Feynman graph factorizes into two disjoint connected graphs after the on-shell propagators are removed. Similar conclusions were drawn in the literature for dimensional-shift identities~\cite{Anastasiou:2013srw,Lee:2012te}. Our result is of course in agreement with reverse unitarity, because inclusive cross sections are discontinuities~\cite{Stapp:1971hh,Polkinghorne:1972hc}, and it extends it to arbitrary cut integrals.

Finally, we note that our result on the difference between leading singularities and maximal cuts is also in agreement with the literature, where it was observed that while leading singularities of Feynman integrals do not necessarily satisfy the IBP identities, some specific linear combinations of leading singularities do (see, e.g., ref.~\cite{Kosower:2011ty,CaronHuot:2012ab,Sogaard:2014jla,Ita:2015tya}). Indeed, there is no contradiction between the fact that cut integrals (including those associated to Landau singularities of the second type) satisfy IBP identities, while individual leading singularities do not. The former compute discontinuities, while the latter do not. Discontinuities and cut integrals may however be sums of leading singularities (cf. the discussion in Section~\ref{sec:leading_singularities}), and so there may be specific linear combinations of leading singularities that satisfy IBP identities.

\subsection{Linear relations among cut integrals in integer dimensions}
\label{sec:relDCI}

In this section we study linear relations among cuts of the integrals in our basis with an even number of propagators, computed in integer dimensions.
Consider the limit $\eps\to0$ of the the differential form defined in eq.~\eqref{eq:compactified},
\begin{equation}\label{eq:comp2m}
	\varpi_{2m}= \frac{d^{2m+2}Y}{\pi^{m}\,\textrm{Vol}(GL(1))}\frac{\delta((YY))}{[-2(X_1Y)]\ldots[-2(X_{2m}Y)]}\,,
\end{equation}
The corresponding cut integrals are
\begin{equation}\label{eq:tildeK_def}
	\cutRes_C\tildeK_{2m}\equiv\lim_{\eps\to0}\cutRes_C\tildeJ_{2m}\,,
	\qquad \emptyset \subset C\subseteq [2m]\,.
\end{equation}
While $\varpi_{2m}$ is always well defined, it might be that for some sets $C$ the corresponding cut integral is not. Indeed, it can happen that when some kinematic scales are set to zero the integral develops poles in $\eps$, or that the limit of sending the scales to zero does not commute with sending $\eps\to 0$.
Both of these cases are excluded from our discussion.

From \refE{eq:comp2m}, we see that $\varpi_{2m}$ is nonsingular for $(X_{\infty}Y)=0$, and so \refE{eq:cut_infty} implies that all cut integrals associated to singularities of the second type must vanish,
\beq
\cC_{\infty C}\tildeK_{2m} = 0\,,\qquad \emptyset \subset C\subseteq [2m]\,.
\eeq
Equation~\eqref{eq:homo_odd} then implies a linear relation among the cut integrals associated to singularities of the first type,
\beq\label{eq:DCI_cut_relations_app}
2\,\cC_C\tildeK_{2m} + \sum_{i\in [2m] \setminus C}\,\cC_{Ci}\tildeK_{2m}= 0\,,\qquad
\textrm{$\emptyset \subset C \subset [2m]$ and $c$ is odd.}
\eeq
In Appendix~\ref{app:Integer_relation} we present an alternative proof which does not rely on homological methods. 
A particular case of this relation for box integrals can be found in ref.~\cite{PhamInTeplitz}.
A relation similar to eq.~\eqref{eq:DCI_cut_relations_app} can be derived from eq.~\eqref{eq:homo_even} for $c$ even, but it is not independent from eq.~\eqref{eq:DCI_cut_relations_app}. Indeed, if $c$ is even, then we can write
\beq\bsp
0&\,=\frac{1}{2}\sum_{j\in [2m]- C}\left[2\,\cC_{C j}\tildeK_{2m} + \sum_{i\in [2m] - C}\,\cC_{C ji}\tildeK_{2m}\right]\\
&\,=\sum_{j\in [2m]- C}\cC_{C j}\tildeK_{2m} + \sum_{\substack{i,j\in [2m] - C\\ i<j}}\,\cC_{C ji}\tildeK_{2m}\,.
\esp\eeq
 We stress that eq.~\eqref{eq:DCI_cut_relations_app}  holds only for integrals with an even number of propagators in integer dimensions. Indeed, if the number $n$ of propagators is odd, then the integral $\tildeK_n$ has a pole on $P_{\infty}$, and the cut integrals associated to singularities of the second type do not vanish. We have checked in explicit examples that eq.~\eqref{eq:DCI_cut_relations_app} does not hold away from integer dimensions.
In the specific case where $n$ is even and $C$ corresponds to all but one propagator, eq.~\eqref{eq:DCI_cut_relations_app} reduces to the relation between the maximal and next-to-maximal cuts, eq.~\eqref{eq:Max-NMax-2_abs}.

\subsection{A linear relation between single and double cuts}
In the previous subsection we obtained relations among cut integrals associated to singularities of the first type by equating two ways to compute the cut integrals associated to the singularities of the second type, eqs.~\eqref{eq:homo_even}, \eqref{eq:homo_odd} and~\eqref{eq:cut_infty}, and using the fact that these integrals have no singularities of the second type. The resulting relations are therefore only valid in the limit $\eps\to0$. In this subsection, we present a relation where the cut integral associated to the singularity of the second type does not vanish, but the integral in eq.~\eqref{eq:cut_infty} is simple enough that it can be performed in closed form, to all orders in dimensional regularization. 

We start by equating eq.~\eqref{eq:homo_even} and~\eqref{eq:cut_infty} for $C=\emptyset$,
\beq\label{eq:inproof}
\sum_{i\in [n]} \cC_{i}I_n + \sum_{\substack{i,j\in [n]\\ i<j}}\cC_{ij}I_n = -2\eps\,\int_{\tilde E_\infty}\varpi_n^D\mod i\pi\,.
\eeq
The integration is performed over the vanishing cell $\tilde  E_{\infty}$, defined as one of the two parts of the quadric $\Sigma$ cut out by the hyperplane $P_{\infty}$. We now show that the value of this integral is half the original Feynman integral,
\beq
\int_{\tilde E_\infty}\varpi_n^D = \frac{1}{2}\int_{\Sigma}\varpi^D_n\mod i\pi\,.
\eeq
Indeed, let us choose coordinates such that $P_{\infty}$ is the hyperplane defined by $Y^+=0$. Then we can decompose $\Sigma = \Sigma^+\cup \tilde S_{\infty} \cup \Sigma^-$, where $\Sigma^\pm = \{Y\in \Sigma: \pm Y^+>0\}$. The vanishing cell can then be identified with one of these two parts, and we may choose without loss of generality $\tilde E_{\infty} = \Sigma^+$.  The integral over $\tilde S_{\infty}$ vanishes (because we integrate the $D$-form $\varpi_n^D$ over the $(D-1)$ sphere $\tilde S_{\infty}$), and so we get the identity
\beq
\int_{\Sigma}\varpi^D_n = \int_{\Sigma^+}\varpi^D_n + \int_{\Sigma^-}\varpi^D_n\,.
\eeq
 Let us now show that the two integrals give identical contributions. Since $\Sigma$ is defined by $(YY)=0$, whenever $Y\in \Sigma^+$ we have $-Y\in \Sigma^-$. The integrand $\varpi_n^D$ is invariant under the change of variables $Y\to -Y$ (up to terms proportional to $i\pi$),
 \beq
\varpi_n^D\to (-1)^{n+x_n+2\eps} \varpi_n^D = (-1)^{n+x_n} \varpi_n^D = \varpi_n^D\mod i\pi\,,
\eeq
because $n+x_n$ is always even ($x_n$ is defined in \refE{eq:xn_def}).
Hence we find
\beq
\int_{\tilde E_{\infty}}\varpi^D_n = \int_{\Sigma^+}\varpi^D_n = \int_{\Sigma^-}\varpi^D_n\,.
\eeq
Putting everything together, we find the following remarkable relation relating the sum of all single and double cuts with the original Feynman integral,
\beq\label{eq:pole_cancellation}
\sum_{i\in [n]} \cC_{i}I_n + \sum_{\substack{i,j\in [n]\\ i<j}}\cC_{ij}I_n = -\eps\, I_n\mod i\pi\,.
\eeq
The special case of this relation for the one-mass box was considered in ref.~\cite{Caron-Huot:2016cwu}.
In Appendix~\ref{app:pole_cancellation} we provide an alternative, purely analytic, proof of eq.~\eqref{eq:pole_cancellation}. The proof presented here relies crucially on the fact that the integral over the vanishing cell in the left-hand side of eq.~\eqref{eq:inproof} is simple enough that it can be performed in closed form. If $C\neq\emptyset$, then the structure of the vanishing cell $\tilde E_{\infty C}$ is more complicated, and we do not currently know how to perform the integral in closed form. We have performed an extensive search  for explicit relations similar to eq.~\eqref{eq:pole_cancellation} for $C\neq\emptyset$, but no such relation was found. 

%%%%%%%%%%%%%%%%%%%%%%%%%%%%%%%%%%%%%%%%%%%%%%

\subsection{A basis of one-loop cut integrals}
\label{sec:cut_masters}

It is well known that using IBP identities every (scalar) Feynman integral can be written as a linear combination of a minimal set of integrals called \emph{master integrals}. It is known that the number of master integrals is always finite~\cite{Smirnov:2010hn}, related to a sum of Milnor numbers of critical points~\cite{Lee:2013hzt}. At one loop, we can choose the basis integrals to lie in different dimensions, and one finds that the integrals $\tildeJ_n$ form a basis of all one-loop integrals in even dimensions.\footnote{We keep in mind that some two- and three-point functions are reducible, and are therefore not independent basis elements.}
Since cut integrals satisfy the same IBP identities as their uncut analogues, it is natural to ask if we can write down a basis for one-loop cut integrals. 
In the following we discuss two such bases associated to singularities of the first and second types.

Since cut and uncut integrals satisfy the same IBP and dimensional shift identities, we immediately conclude that $\{\cC_C\tildeJ_n: \emptyset\subset C\subseteq [n]\cup\{\infty\}\}$ is a spanning set of all one-loop integrals. This set, however, is not yet minimal, because cut integrals satisfy additional linear relations coming from relations among the generators of the homology groups (cf. Section~\ref{sec:cut_infty_rel}). 
In particular, eqs.~\eqref{eq:homo_even} and~\eqref{eq:homo_odd} imply that we can always express cut integrals associated to singularities of the second type in terms of those of the first type, and so we conclude that the cut integrals 
\beq\label{eq:basis_1}
\{\cC_C\tildeJ_n: \emptyset\subset C\subseteq [n]\} 
\eeq 
associated to singularities of the first type 
suffice to form a spanning set of all one-loop cut integrals. We think of this set as a basis, although strictly speaking, just as for Feynman integrals,  some two- and three-point integrals are reducible. This basis has the property that all the basis elements are polylogarithmic functions of uniform weight $\lceil n/2\rceil - \lceil c/2\rceil$, order by order in dimensional regularization. In particular, the maximal cut is a function of weight 0, i.e., an algebraic function at $\eps=0$ (cf.~the discussion on leading singularities of Section~\ref{sec:leading_singularities}).

There is a natural alternative basis for one-loop cut integrals, which mixes cut integrals associated to singularities of the first and second types. Indeed, we can use eq.~\eqref{eq:homo_odd} and replace each cut integral $\cC_C\tildeJ_n$, $c$ odd, by the cut integral $\cC_{\infty C}\tildeJ_n$. The alternative basis is~\cite{PhamInTeplitz}
\beq\label{eq:basis_2}
\{\cC_C\tildeJ_n: \emptyset\subset C\subseteq [n]\cup\{\infty\} \textrm{ and } c \textrm{ even}\}\,. 
\eeq 

We  conclude this section with a comment on the differential equations satisfied by cut integrals. It is well known that (uncut) master integrals satisfy systems of first-order linear differential equations~\cite{Kotikov:1990kg,Kotikov:1991hm,Kotikov:1991pm,Gehrmann:1999as,Henn:2013pwa}. The differential equations are obtained by differentiating under the integration sign. The differentiation generically introduces integrals with higher powers of the propagators, which may be reduced to a linear combination of master integrals. Since uncut and cut integrals satisfy the same IBP identities, we immediately conclude that cut integrals satisfy the same differential equations as uncut Feynman integrals. We stress that this argument is independent of the loop order, and it agrees with the reverse-unitarity approach to the computation of inclusive cross sections~\cite{Anastasiou:2002yz,Anastasiou:2002wq,Anastasiou:2002qz,Anastasiou:2003yy,Anastasiou:2003ds} and recent approaches to solve homogeneous differential equations by maximal cuts~\cite{Remiddi:2016gno,Primo:2016ebd}. As a consequence, the bases of one-loop cut integrals in eq.~\eqref{eq:basis_1} and~\eqref{eq:basis_2} satisfy the same differential equations as the corresponding integrals $\tildeJ_n$.

% !TEX root = main-cuts.tex

\section{Conclusions}
\label{sec:conclusions}

In this paper, we have presented a definition of one-loop cut integrals valid in dimensional regularization, for any configuration of internal or external scales. The cornerstone of our definition is Leray's multivariate residue calculus, which allows us to make a precise definition for the sometimes rather vague notion of the integration contour over which a cut integral should be evaluated. Leray's multivariate residue calculus is intimately connected to homology theory, and we have studied the homology groups associated to one-loop integrals. The study of the homology groups gives us precise relations between cut integrals associated to Landau singularities of the first and second types and discontinuities around the corresponding Landau varieties. Moreover, relations between the generators of the homology groups translate immediately into linear relations between cut integrals with different numbers of cut propagators.

In analyzing the consequences of the homology relations, we have used a representation of one-loop integrals in a compactified space. In this framework, cuts relating to Landau singularities of the first and second kind are described together,  where the only difference between them is that the latter involves cutting a special propagator which is raised to a non-integer power. 
Furthermore, in the compactified framework we have identified a single class of parametric integrals that yields generic uncut and cut one-loop integrals as special cases, hence establishing a new way to evaluate cut integrals (see eqs.~\eqref{eq:cut-loop-duality} and (\ref{eq:paramIntai})) without explicitly parametrizing momenta.

While the work in this paper was restricted to the study of cuts of one-loop integrals, we believe that several of the concepts introduced in this paper carry over to higher loop integrals. In particular, the theory of multivariate residues is not restricted to one loop, and it was already realized in the '60s that it provides the natural language to study cuts and discontinuities of Feynman integrals~\cite{Froissart,PhamBook,PhamCompact,PhamInTeplitz,Teplitz,Bloch:2015efx}. The study of the homology groups associated to multi-loop integrals, however, is much more complicated, and only very limited results are available in the literature~\cite{Federbush}. The study of the homology groups at one-loop was crucial in order to define a basis of integration contours for one-loop integrals, and we believe that the homology groups of higher loop integrals can provide new insight into the structure of higher loop integrals. A possible avenue for future research is the application of homology theory to construct so-called master contours. Beyond one-loop the construction of master contours, i.e., integration contours which allow one to project an amplitude onto a given master integral, is still a largely open question. The current approach to the construction of master contours relies on the computation of leading singularities, and the projectors correspond to linear combinations of leading singularities that preserve IBP identities. Since the generators of the homology groups allow one to compute discontinuities, they naturally provide a basis of integration contours that preserve IBP relations. The exploration of a possible connection between the generators of the homology groups and master contours is left for future work.

\acknowledgments
We are grateful to Harald Ita, Roman Lee, Ben Page, Erik Panzer and Volodya Smirnov for discussions and communications.
SA acknowledges the hospitality of Trinity College Dublin and the CERN Theoretical Physics Department at various stages of this work.
CD acknowledges the hospitality of the Higgs Center of the University of Edinburgh and of Trinity College Dublin at various stages of this work. EG acknowledges the hospitality of Trinity College Dublin.  
This work is supported by the Juniorprofessor Program of Ministry of Science, Research and the Arts of the state of Baden-W\"urttemberg, Germany (SA), the ERC Consolidator Grant 647356 ``CutLoops'' (RB), the ERC Starting Grant 637019 ``MathAm'' (CD), and by the STFC Consolidated Grant ``Particle Physics at the Higgs Centre'' (EG). 
We would also like to thank the ESI institute in Vienna and the organizers of the program ``Challenges and Concepts for Field Theory and Applications in the Era of LHC Run-2'', and Nordita in Stockholm and the organizers of ``Aspects of Amplitudes,'' over the summer of 2016, where certain ideas presented in this paper were consolidated.

%%%%%%%%%%%%%%%%%%%%%%%%%%%%%%%%%%%%%%
%%%%%%%%%%%%%%%%%%%%%%%%%%%%%%%%%%%%%%
%%%%%%%%%%%%%%%%%%%%%%%%%%%%%%%%%%%%%%

%						APPENDICES

%%%%%%%%%%%%%%%%%%%%%%%%%%%%%%%%%%%%%%
%%%%%%%%%%%%%%%%%%%%%%%%%%%%%%%%%%%%%%
%%%%%%%%%%%%%%%%%%%%%%%%%%%%%%%%%%%%%%

 \appendix
 % !TEX root = main-cuts.tex

\section{Changing variables to the propagators}
\label{app:Baikov}

In this appendix we derive eq.~\eqref{eq:Baikov} (see also ref.~\cite{Frellesvig:2017aai,Baikov:1996iu,Lee:2010wea,Grozin:2011mt}). For convenience, we only discuss the Euclidean case. The extension to Minkowski space is straightforward. We use the notation of Section~\ref{sec:def_cuts}. We can write $k^E = k_{\parallel} + k_{\bot}$, and so
\beq
d^Dk^E = d^{c-1}k_{\parallel}\,d^{D-c+1}k_{\bot} = \frac{1}{2}\,d^{c-1}k_{\parallel}\,d\Omega_{D-c}\,(k_{\bot}^2)^{(D-c-1)/2}\,dk_{\bot}^2\,.
\eeq
In the linear subspace $\cE_C$ we change variables to $\kappa_j\equiv k^E\cdot (q^E_j-q^E_{1})$, $2\le j\le c$. The jacobian is given by the determinant
\beq
J = \det\left(\frac{\partial \kappa_j}{\partial k^E_{\mu}}\right)_{2\le j\le c} = \det\left((q^E_j-q^E_{1})^\mu\right)_{2\le j\le c}\,.
\eeq
It is easy to check that $J$ is the square root of the Gram determinant, $J^2= \det((q^E_j-q^E_{1})\cdot(q^E_l-q^E_{1}))_{2\le j,l\le c} = \Gram_C$. Using the fact that (see Fig.~\ref{fig:simplices}) $|k_{\bot}| = H_C/\Gram_C$, we obtain
\beq\bsp
d^Dk^E &\,= \frac{1}{2\sqrt{H_C}}\,\left(\frac{H_C}{\Gram_C}\right)^{(D-c)/2}\,d^{c-1}\kappa_{j}\,d\Omega_{D-c}\,dk_{\bot}^2\\
&\,= \frac{1}{2\sqrt{H_C}}\,\left(\frac{H_C}{\Gram_C}\right)^{(D-c)/2}\,d^{c-1}\kappa_{j}\,d\Omega_{D-c}\,d(k^E)^2\,,
\esp\eeq
where in the last step we have changed variables from $k_{\bot}^2$ to $(k^E)^2$. Finally, we can change variables from $((k^E)^2,\kappa_j$) to the propagators $D_i$. The change of variables is given by
\beq
D_{1} = (k^E)^2 +m_{1}^2 \,, \qquad D_j = (k^E)^2 -2 k^E\cdot q^E_j + (q_j^E)^2 +m_{j}^2\,,\qquad 2\le j\le c\,.
\eeq
The jacobian is $2^{1-c}$, and so we find
\beq\bsp
d^Dk^E &\,= \frac{2^{-c}}{\sqrt{H_C}}\,\left(\frac{H_C}{\Gram_C}\right)^{(D-c)/2}\,d\Omega_{D-c}\,\prod_{j\in C}dD_j\,,
\esp\eeq
in agreement with eq.~\eqref{eq:Baikov}.
  % !TEX root = main-cuts.tex

\section{The Landau conditions at one loop}
\label{app:Landau}

In this appendix we solve the Landau conditions for one-loop integrals, and we show that they can be classified according to the vanishing of a modified Cayley determinant $Y_C$ or a Gram determinant $\Gram_C$, where $C$ is a subset of propagators.
The results summarized here agree with those obtained after compactification in \refS{sec:compactification}.
It is convenient to work with the Feynman parameter representation integral of one-loop integrals,
\beq
I_n^D = (-1)^n\,e^{\gamma_E\eps}\,\Gamma(n-D/2)\,\int_{\Delta}\Omega_{n-1}\,\frac{\mathcal{U}_n^{n-D}}{\mathcal{F}_n^{n-D/2}}\,,
\eeq
where $\Delta$ is the standard simplex in $\mathbb{RP}^{n-1}$ and $\Omega_{n-1} = \delta\left(1-\sum_{i\in S}x_i\right)\prod_{i=1}^ndx_i$ is the usual volume form on $\mathbb{RP}^{n-1}$ ($S$ is any subset of $\{1,\ldots,n\}$). $\mathcal{U}_n$ and $\mathcal{F}_n$ denote the two Symanzik polynomials for one-loop graphs \cite{Bogner:2010kv},
\beq\bsp
\mathcal{U}_n \, = \sum_{i=1}^nx_i\,,\qquad\qquad
\mathcal{F}_n \, = \sum_{i,j=1}^n\Upsilon_{ij}\,x_i\,x_j\, = \,\vec x^T\,\Upsilon\,\vec x\,,
\esp\eeq
with $\Upsilon_{ij} = \frac{1}{2}(-(q_i-q_j)^2+m_i^2+m_j^2)$. Note that the principal minor of $\Upsilon$ where all rows and columns are deleted except for those in a set $C \subseteq \{1,\ldots,n\}$ is precisely the modified Cayley determinant $Y_C$. 

The Landau conditions~\eqref{eq:Landau_1} and~\eqref{eq:Landau_2} take the form
\beq\bsp\label{eq:Landau_cond_1-loop}
&\alpha_U\,\mathcal{U}_n = 0\,,\qquad \alpha_F\,\mathcal{F}_n = 0\,,\qquad \beta_i\,x_i = 0\,,\qquad 1\le i\le n\,,\\
&\alpha_U \vec{1}+ \alpha_F\,\Upsilon\vec x + \vec\beta = 0\,,
\esp\eeq
with $\vec 1 = (1,\ldots,1)$ and $\vec\beta=(\beta_1,\ldots,\beta_n)$. We are looking for constraints on the external kinematic variables for which the Landau conditions have nontrivial solutions, i.e., solutions for which $(\alpha_U,\alpha_F,\beta_1,\ldots,\beta_n)\neq (0,\ldots,0)$. 

We first analyze the solutions for which a subset of the $\beta_i$ are nonzero. Equation~\eqref{eq:Landau_cond_1-loop} implies that the corresponding subset of Feynman parameters $x_i$ must vanish, and it is easy to check that in that case $\mathcal{U}_n$ and $\mathcal{F}_n$ reduce to the Symanzik polynomials of the one-loop graph where all edges with a vanishing $x_i$ are pinched. In other words, every solution to the Landau conditions of a pinched graph is also a solution for the full graph. It is therefore sufficient to study the \emph{leading Landau singularities} of the graph, defined by $\vec\beta=\vec 0$.

Thus we focus on the leading Landau singularities. It is easy to see that a nontrivial solution must satisfy $(\alpha_U,\alpha_F)\neq(0,0)$. Moreover, eq.~\eqref{eq:Landau_cond_1-loop} implies that for $\alpha_F=0$ we only obtain the trivial solution. Hence, we consider the following two cases:
\begin{enumerate}
\item If $\alpha_U=0$, the Landau conditions reduce to $\Upsilon\vec x=0$, and this system has a nontrivial solution if and only if the modified Cayley determinant of the one-loop graph vanishes, $\det\Upsilon=0$.
\item If $\alpha_U\neq0$, then the Landau conditions reduce to 
\beq\label{eq:Landau_aUaF}
\Upsilon\vec x + \alpha\vec 1=0\,, 
\eeq
with $\alpha \equiv \alpha_U/\alpha_F \neq0$. We already know that we obtain nontrivial solutions to the Landau conditions for $\det\Upsilon=0$. Therefore we only discuss the case where $\Upsilon$ is invertible, in which case the unique solution to eq.~\eqref{eq:Landau_aUaF} is $\vec x = -\alpha\Upsilon^{-1}\vec 1$. This provides a constraint on the  kinematic data, because $\mathcal{F}_n=0$ implies
\beq
0 =  \vec 1^T\Upsilon^{-1}\vec 1 = \sum_{i,j=1}^n (\Upsilon^{-1})_{ij}\,.
\eeq
The previous condition is equivalent to the solutions of the Landau conditions of the second type, which correspond to the vanishing of the Gram determinant. Indeed, it can be shown that
\beq\label{eq:inverse_Y_sum}
\sum_{i,j=1}^n (\Upsilon^{-1})_{ij} = \frac{\Gram_{[n]}}{\det \Upsilon}\,.
\eeq
\end{enumerate}

By lifting these two criteria for leading Landau singularities of pinched graphs to the original unpinched graph, we obtain the two criteria of vanishing $Y_C$ or $\Gram_C$ for a subset $C$ of propagators.
 
 % !TEX root = main-cuts.tex
\section{Analytic proofs of relations among cut integrals}

\subsection{Relations among cut integrals in integer dimensions}
\label{app:Integer_relation}

In this Appendix we present a purely analytic proof of eq.~\eqref{eq:DCI_cut_relations_app} which does not rely on homological methods. We exclude the problematic cases discussed in \refS{sec:relDCI}, and only discuss integrals that are finite when $\eps\to0$.
The proof proceeds by induction on $m$ and relies on
two key properties:  first, that cut operators commute with differential operators (see Section~\ref{sec:cut_masters}), and second, that 
the total differential of these integrals with $2m+2$ propagators can be expressed in terms of the integrals with $2m$ propagators \cite{Spradlin:2011wp,Goncharov:1996}.  
We seed the induction with the case $m=1$, which we can check explicitly:
\begin{align}\bsp\label{eq:cutsbub2m}
 	\cutRes_{e_1e_2}\tildeK_2(p^2;m_1^2,m_2^2)&=\frac{2}{\lambda(m_1^2,m_2^2,p^2)}\,,\\
 	\cutRes_{e_1}\tildeK_2(p^2;m_1^2,m_2^2)&=\cutRes_{e_2}\tildeK_2(p^2;m_1^2,m_2^2)=
 	-\frac{1}{\lambda(m_1^2,m_2^2,p^2)}\,.
\esp\end{align}
These are obtained using eqs.~\eqref{eq:MaxCutEven} and \eqref{eq:NMaxCutEven}.
We assume that \refE{eq:DCI_cut_relations_app} holds for all integers up to a given value of $m$, and then prove it for $m+1$.

Throughout the proof it is convenient to work with integrals normalized as follows:
\beq\label{eq:norm_K}
\plainK_n \equiv \tildeK_n/j_n\,.
\eeq
Now we rely on a recursive differential equation for these integrals given in ref.~\cite{Goncharov:1996,Spradlin:2011wp}\footnote{This formula was given under the assumption of vanishing internal masses, but it is straightforward to extend the relation to finite masses using the compactified picture of Section~\ref{sec:compactification}.},
\beq \label{eq:automotive}
d\plainK_{2m+2} = 
\sum_{\substack{S \subset [{2m+2}] \\ \abs{S}=2m}}
\plainK_{S} \, d \log R_S
\eeq
where $R_S$ is some algebraic function, and $K_{S}$ denotes the function $K_{\abs{S}}$ evaluated at $(X_j)_{j\in S}$.  
Now we can use this differential equation to implement the inductive step from $2m$ to $2m+2$.  Let us take the total differential of the left-hand side of \refE{eq:DCI_cut_relations_app}.  We have
\beq\bsp
d& \left[2\,\cC_C
 + \sum_{i\in [{2m+2}] \setminus C}\,\cC_{C i} \right]
 \plainK_{{2m+2}}
\\
 &=  \sum_{\substack{S \subset [{2m+2}] \\ \abs{S}=2m}}
 \left[2\,\cC_C \plainK_S
 + \sum_{i\in [{2m+2}] \setminus C}\,\cC_{C i} \plainK_S\right] \, d \log R_S
 \\
 &= 0\,,
 \esp
\eeq
where the last line follows from the induction hypothesis.  We conclude the right-hand side of \refE{eq:DCI_cut_relations_app} must be a constant.  The constant can only be 0, because the signs of the integrals $K_S$ can be reversed simply by changing the normalization convention (the sign of the square root \eqref{eq:LS_J_n}).  This inductive argument takes care of all cases with $|C|<2m$.  The only remaining possibility is $|C|=2m-1$, which is simply the statement of \refE{eq:Max-NMax-2_abs}.

 % !TEX root = main-cuts.tex

\subsection{Relation among cut and uncut integrals}\label{app:pole_cancellation}

We now present a purely analytic proof of \refE{eq:pole_cancellation}.
Since this is a linear relation, it suffices to prove it for elements of the normalized basis $\plainJ_n$ defined as:
\begin{equation}
	\plainJ_n=\tildeJ_n/j_n\,.
\end{equation}
For compactness of the equations below, we define
\begin{equation}\label{eq:cancPoleApp}
	Q_n\equiv\sum_{i=1}^{n}\cC_{i}\plainJ_n+\sum_{\substack{i,j=1\\ i< j}}^{n} \cC_{ij}\plainJ_n + \eps\,\plainJ_n\,,
\end{equation}
and similarly for $\widetilde Q_n$ defined in terms of the $\tildeJ_n$. Our goal is to show that $Q_n=0$ for all $n$.

We proceed by induction on $n$.  It is to be understood that  all identities in this derivation are valid modulo $i\pi$.\footnote{It is possible that a precise prescription for integration contours would make the relations valid exactly. We  do not yet see an obvious choice of contour, but this is an interesting avenue for further exploration.}
It is simple to check that the relation holds for the tadpole,
\beq
Q_1=\cC_1\plainJ_1 + \eps \,\plainJ_1 = 0\,.
\eeq

Let us now assume that $Q_{j}=0$  up to $j=n-1$, and let us check that it still holds for $j=n$. Since the $\plainJ_n$ form a basis of one-loop integrals, for every $\plainJ_n$ we can write a differential equation of the following form (cf.~\refS{sec:cut_masters}):
\beq
d\plainJ_n = \sum_{\substack{I\subseteq [n]\\I\neq \emptyset}}A_n^I\,\plainJ_{I}\,,
\eeq
where $\plainJ_{I}$ denotes the integral in which all the propagators of $\plainJ_n$ have been contracted except for those in the subset $I$ and the $A_n^I$ are algebraic functions of the kinematic invariants and $\eps$. All cuts of $\plainJ_n$ satisfy the same differential equation.  In particular, we note the relations in which one, two, or $n$ propagators of $\plainJ_n$ have been cut:
\begin{align}\bsp
d\cC_i\plainJ_n =& \sum_{\substack{I\subseteq [n]\\i\in I}}A_n^I\,\cC_i\plainJ_{I}\,,\qquad
d\cC_{ij}\plainJ_n = \sum_{\substack{I\subseteq [n]\\i,j\in I}}A_n^I\,\cC_{ij}\plainJ_{I}\, ,\qquad
d\cC_n\plainJ_n =  A_n^{[n]}\,\cC_n\plainJ_{n}\,.
\label{eq:dencut}
\esp\end{align}
We recall that $\cC_i\plainJ_{I}=0$ unless $i\in I$, and similarly for $\cC_{ij}\plainJ_{I}$.

With the  relations in \refE{eq:dencut}, consider the action of the differential on $Q_n$.  We find
\beq\bsp
dQ_n\,=\sum_{\substack{I\subseteq [n]\\ I\neq \emptyset}}A_n^I\,\left[\sum_{1\le i\le n}\cC_{i}\plainJ_{I}+\sum_{1\le i<j\le n} \cC_{ij}\plainJ_{I} + \eps\, \plainJ_{I}\right]\,.
\esp\eeq
By the induction hypothesis, all the terms vanish except for the homogenous term,
\beq\bsp
dQ_n\, = A_n^{[n]}\,Q_n\,.
\esp\eeq
The coefficient $A_n^{[n]}$ is completely determined by \refE{eq:dencut} and the formula for the maximal cut given in
eq.~\eqref{eq:MaxCutEven} and~\eqref{eq:MaxCutOdd}.  The result is
\beq
A_n^{[n]} = -\eps\,d\log \left(\frac{Y_n}{\Gram_n}\right)\,.
\eeq
Hence, we obtain a relation of the form
\beq\label{eq:ansatz_pole_cancellation}
Q_n = C_n(\eps)\,\left(\frac{Y_n}{\Gram_n}\right)^{-\eps}\,,
\eeq
where the function $C_n(\eps)$ is not yet determined.  In order to finish the proof, we need to show that $C_n(\eps)=0$ for all $n$.  We do this by analyzing the behavior of eq.~\eqref{eq:ansatz_pole_cancellation} in the soft limit where two internal momenta coincide, $q_a=q_b$.  We will do this in the generic case where all masses are different (in particular $m_a^2\neq m_b^2$). The non-generic case then follows as a limit of the generic case (we work to all orders in $\eps$ in dimensional regularization, so the massless limit of eq.~\eqref{eq:ansatz_pole_cancellation} is smooth before expansion in $\epsilon$).

We multiply both sides of eq.~\eqref{eq:ansatz_pole_cancellation} by  ${j}_n$, giving
\beq\label{eq:ansatz_pole_cancellation_LS}
C_n(\eps) = \frac{1}{{j}_n}\,\left(\frac{Y_n}{\Gram_n}\right)^{\eps}\,\widetilde Q_n\,.
\eeq
Since the integral converges, we can take the limit under the integration for the different terms in $\widetilde Q_n$ and use the partial fraction identity
\beq
\frac{1}{((k-q_a)^2+m_a^2)\,((k-q_a)^2+m_b^2)} = \frac{1}{m_a^2-m_b^2}\,\left(\frac{1}{(k-q_a)^2+m_b^2} - \frac{1}{(k-q_a)^2+m_a^2}\right)\,.
\eeq
We obtain
\beq\bsp
\lim_{q_b\to q_a}\tildeJ_n &\,= \frac{1}{m_a^2-m_b^2}\,\left(\tildeJ_{n-1}^{(a)}-\tildeJ_{n-1}^{(b)}\right)\,,
\\
\lim_{q_b\to q_a}\cC_i\tildeJ_n &\,= \frac{1}{m_a^2-m_b^2}\,\left(\cC_i\tildeJ_{n-1}^{(a)}-\cC_i\tildeJ_{n-1}^{(b)}\right)\,,\\
\lim_{q_b\to q_a}\cC_{ij}\tildeJ_n &\,= \frac{1}{m_a^2-m_b^2}\,\left(\cC_{ij}\tildeJ_{n-1}^{(a)}-\cC_{ij}\tildeJ_{n-1}^{(b)}\right)\,,
\esp\eeq
where $\tildeJ_{n-1}^{(a)}$ denotes the integral obtained from $\tildeJ_n$ by removing the propagator of momentum $q_a$.
In these expressions, the cut of an absent propagator (for example $\cC_a\tildeJ_{n-1}^{(a)}$) is identically  zero.
Since the induction hypothesis implies that \refE{eq:cancPoleApp} is satisfied by $\tildeJ_{n-1}^{(a)}$ and $\tildeJ_{n-1}^{(b)}$,
we  conclude
 that
\beq\bsp
\lim_{q_b\to q_a}\widetilde Q_n=0\, ,
\esp\eeq
which completes our argument.

\bibliographystyle{JHEP}
\bibliography{bibMain.bib}

\end{document}